\documentclass[aps,prd,amssymb,eqsecnum,12pt,preprint]{revtex4}
\usepackage{graphicx,color}
\usepackage{bm,amsmath}

\begin{document}

\newcommand{\sigmap}{\sigma_\mathrm{p}}
\newcommand{\Zp}{Z_\mathrm{p}}
\newcommand{\sgn}{\mathrm{sign}}
\newcommand{\braket}[1]{{\mathinner{\langle{#1}\rangle}}}

\vspace*{1cm}
\title{
\Large
Perturbations of Spacetime around a Stationary Rotating Cosmic String 
\bigskip
}

\author{ \large 
Kouji Ogawa\footnote{Present address: 
Observations Division,  
Fukuoka District Meteorological Observatory, Fukuoka.}, 
 Hideki Ishihara\footnote{
E-mail:ishihara@sci.osaka-cu.ac.jp}, 
 Hiroshi Kozaki$^{1}$, and 
 Hiroyuki Nakano$^2$
\bigskip
}
\affiliation{
Department of Mathematics and Physics,~Graduate School of Science, 
Osaka City University, Osaka 558-8585, Japan
\bigskip
\\
{}$^1$
Department of General Education,
Ishikawa National College of Technology,
Tsubata, Kahoku-gun, Ishikawa 929-0392, Japan
\bigskip
\\ 
{}$^2$Center for Computational Relativity and Gravitation, 
School of Mathematical Sciences, 
Rochester Institute of Technology, 
Rochester, New York 14623, USA
\bigskip
}

\begin{abstract}
We consider the metric perturbations around a stationary rotating 
Nambu-Goto string in Minkowski spacetime.
By solving the linearized Einstein equations, we study 
the effects of azimuthal frame-dragging around the rotation axis 
and linear frame-dragging along the rotation axis, 
the Newtonian logarithmic potential, 
and the angular deficit around the string as the potential mode. 
We also investigate gravitational waves propagating off the string and 
propagating along the string, and 
show that the stationary rotating string 
emits gravitational waves toward the directions specified by discrete 
angles from the rotation axis. 
Waveforms, polarizations, and amplitudes which depend on the direction 
are shown explicitly. 
\end{abstract}

\preprint{OCU-PHYS 307}
\preprint{AP-GR 64}

\date{\today\vspace{2cm}}

\maketitle

\section{Introduction} 
The phase transition of vacuum in the early universe 
is one of the most important topics of cosmology 
and elementary particle physics. 
It is well known that topological defects 
are necessarily created due to the spontaneous symmetry breaking 
of vacuum states\cite{Kibble76} 
(see also \cite{Hindmarsh:1994re, VandS, Anderson}). 
Among the several types of topological defects, 
cosmic strings are possible to survive until the present stage of the universe 
and to be observed by the gravitational effects. 
Alternatively, it is pointed out that fundamental strings and/or D-strings 
can play a role of cosmic strings\cite{Sarangi:2002yt, Jones:2003da, 
Dvali:2003zj, Copeland:2003bj, JJP}.  
There is no doubt that detection of cosmic strings in the present stage of the 
Universe is important and challenging work.

The gravitational waves from cosmic strings is one of the targets 
of ongoing experiments for searching gravitational waves 
due to recent technological advance, e.g., LIGO, LISA, VIRGO, 
TAMA300, GEO600 and so on\cite{LIGO,LISA,VIRGO,TAMA,GEO}, and 
also theoretical research has been established. 
For example, there are many works on 
the gravitational waves produced by oscillating loop 
cosmic strings\cite{Vilenkin:1981bx, Barden_GW}, 
by an infinitely long string with a helicoidal standing wave\cite{Maria}, 
and by colliding wiggles on a straight string\cite{Hindmarsh, Siemens:2001dx}. 
Damour and Vilenkin\cite{DandV, DandV2} discussed the gravitational wave 
bursts from cusps of the cosmic string. 

A conical spacetime generated around a straight string makes undistorted 
double images of a distant source. The gravitational lensing caused by 
the cosmic strings is studied extensively\cite{lensing}. 
Recently, a variety of gravitational lensing: weak lensing\cite{Kuijuken}, 
lensing by string loops\cite{Mack}, 
and lensing by strings with small-scale structure,\cite{Dyda} was studied. 

It is known that reconnection probability  for gauge theory strings is 
essentially 1\cite{ShellardCSInt}. 
Such the strings 
evolve in a scale invariant way (see \cite{VandS} and references therein). 
In contrast, regarding 
the cosmic strings in the framework of the superstring theory, 
the reconnection probability is suppressed 
sufficiently $<1$~\cite{Jones:2003da, Dvali:2003zj, Copeland:2003bj, JJP}. 
Evolution of such strings may differ from that of gauge strings. 
If the strings are practically stable, we could expect that they 
survive finally in the stationary states 
in the present stage of the Universe.

Starting from the pioneering work by Burden and Tassie\cite{BT},
there are many works on the stationary rotating strings\cite{Frolov, VLS}.  
In our previous study~\cite{RRS}, we reformulate the stationary rotating 
strings as an example of the cohomogeneity-one strings\cite{IshiharaKozaki,KKI}. 
Because of the geometrical symmetry of the strings, 
it is easy to treat them as gravitational sources in the frame work of 
general relativity. 
In this paper, we investigate the gravitational fields around 
a stationary rotating string by solving the linearized Einstein equations 
toward detection of the strings in the universe. 
The Newtonian logarithmic potential and angular deficit are obtained 
as the potential mode. 
Furthermore, two effects of frame-dragging are shown: 
azimuthal dragging around the rotation axis, 
and linear dragging along the rotation axis. 
We also study the gravitational waves propagating 
off the strings and propagating along the strings (traveling waves). 
Characteristic properties of waveforms, polarization, and directions of 
emission are discussed. 

This paper is organized as follows. 
In Sec. II, we briefly review the stationary rotating strings following 
Ref.\cite{RRS}. 
In Sec.III, we formulate linear perturbations of the metric around a 
stationary rotating string. 
We obtain solutions to the linearized Einstein equations explicitly 
then discuss a potential mode in Sec.IV,   
and the gravitational wave modes in Sec.V. 
The traveling wave modes are discussed in Sec.VI. 
Finally, we summarize in Sec.VII. 
We use the sign convention $-\,+\,+\,+$ for the metric, 
and units in which $c=G=1$.

\section{Solutions of stationary rotating strings} 
\subsection{Stationary rotating Nambu-Goto strings in Minkowski spacetime} 
We consider cosmic strings which are 
described by the Nambu-Goto action, 
\begin{align}
	S_\mathrm{NG} = -\mu \int_\Sigma d^2 \zeta \ \sqrt{- \gamma }, 
\label{NG} 
\end{align}
where $\Sigma$ is a timelike two-dimensional world surface 
embedded in a target spacetime $\cal M$ with the metric $g_{\mu\nu}$, 
$\zeta^a ~(\zeta^0=\tau, \zeta^1=\sigma)$ are coordinates on $\Sigma$, 
$\gamma$ is the determinant of the induced metric $\gamma_{ab}$  on $\Sigma$, 
and a constant $\mu $ denotes the string tension. 
Varying the action \eqref{NG} by the coordinates of $\cal M$, 
$ x^\mu~(\mu= 0, 1, 2, 3)$, we obtain the Nambu-Goto equations:
\begin{align}
	\frac{1}{\sqrt{- \gamma }} \ 
	\partial_a\left( \sqrt{- \gamma } \gamma^{ab} \partial_b x^\mu \right)
	+ \Gamma^\mu_{\nu\lambda } 
	\gamma^{ab} \partial_a x^\nu \partial_b x^\lambda =0 , 
\label{NGeq} 
\end{align}
where $ \Gamma^\mu_{\nu\lambda }$ is the Christoffel symbol associated 
with $g_{\mu \nu}$.

When the world surface of a string $\Sigma$ is tangent to 
a Killing vector field in a target spacetime $\cal M$, 
i.e., cohomogeneity-one string, the Nambu-Goto equation \eqref{NGeq} 
can be reduced to a geodesic equation 
in an appropriate three-dimensional metric\cite{Frolov, IshiharaKozaki, RRS}. 
Here, we concentrate on stationary rotating strings, which belong to a class of
the cohomogeneity-one strings. 
We briefly review the solutions of 
stationary rotating strings in Minkowski spacetime according to \cite{RRS}.

In Minkowski spacetime 
with the metric by the cylindrical coordinate system, 
\begin{align}
	ds^2 = g_{\mu \nu} dx^\mu dx^\nu 
		= -dt^2 + d\rho^2 + \rho^2 d\varphi^2 +dz^2,
\label{MinMetCyl}
\end{align}
the Killing vector field $\xi$ which describes the stationary rotation 
around the $z$ axis with a constant angular velocity $\Omega$ is 
\begin{equation}
\begin{aligned}
	{\xi}= \partial_t + \Omega \partial_\varphi.
\end{aligned}\end{equation}
We consider a world surface $\Sigma$ of a stationary rotating string 
which is tangent to $\xi$. 
The solutions are characterized by two dimensionless parameters 
$ l $ and $q$, and explicit forms are given by
\begin{equation}
\begin{aligned}
	&t (\tau ) = \tau ,
\\
	&{\rho}(\sigma)^2 
		= \frac{1}{2} \left\{
			(\rho_\mathrm{max}^2+\rho_\mathrm{min}^2) 
				-(\rho_\mathrm{max}^2-\rho_\mathrm{min}^2)
			\cos\left(2\Omega \sigma \right) 
		\right\} , 
\\ 
	&\varphi(\tau, \sigma) = \Omega \tau + \bar{\varphi } (\sigma) ,  
\\
	&z (\sigma ) = q \sigma , 
\end{aligned}
\label{SRSsol}
\end{equation}
where $\bar\varphi(\sigma)$ is implicitly given by 
\begin{align}
	&\frac{2 l}{\Omega^2}\tan\left(\bar{\varphi}(\sigma) - \varphi_0 
		+  l  |\Omega| \sigma \right)
		= 	
	(\rho_\mathrm{max}^2+\rho_\mathrm{min}^2)
		\tan\left(|\Omega| \sigma+\frac{\pi}{4} \right) 
		- (\rho_\mathrm{max}^2-\rho_\mathrm{min}^2),
\label{sol_phi} 
\end{align}
and $\rho_\mathrm{min}$, $\rho_\mathrm{max}$ are defined by
\begin{equation}\begin{aligned}
	& \rho_\mathrm{min}^2 = \frac{1}{2\Omega^2}\left(
			1+ l^2-q^2 
	- \sqrt{(1+ l +q)(1+ l  -q)(1- l  +q)(1- l  -q)}\right), 
\\
	& \rho_\mathrm{max}^2 = \frac{1}{2\Omega^2}\left(
			1+ l ^2-q^2
	+ \sqrt{(1+ l  +q)(1+ l  -q)(1- l  +q)(1- l  -q)}\right).
\label{minmax}
\end{aligned}\end{equation}
The constant $\varphi_0$ has been fixed for convenience as 
\begin{equation}
	\tan \varphi_0 = -\frac{\Omega^2 \rho^2_\mathrm{min}}{l}, 
\label{phi_0}
\end{equation}
in order that $\bar\varphi=0$ when $\sigma=0.$

The range of $ l $ and $q$ are limited for the stationary rotating strings 
as
\begin{align}
	| l | + |q| \le 1. 
\label{lpCond0}
\end{align}
We do not consider the case $q=0$ in which the Killing vector $\xi$ 
becomes null at the end points of the stationary string.
Changes of sign of parameters $ l , q$ and $\Omega$ can be interpreted as 
reflection of the space and time. Then, we consider, hereafter, the case 
\begin{equation}
	 l \geq 0,\quad q>0,\quad \Omega>0,\quad \mbox{and}\quad  l  + q \le 1.
\end{equation}


In the stationary rotating string solutions 
\eqref{SRSsol}-\eqref{phi_0}, we use the parameters $\tau$ and $\sigma$ which 
respect the Killing vector $\xi$, i.e., 
$	\xi=\frac{\partial}{\partial\tau}. 
$
In contrast, using the conformal flat gauge, which is normally used, 
Burden gave a clear expression for the solutions\cite{Burden:2008zz}.

\begin{figure}[ht]
\begin{center}
\includegraphics[width=10cm]{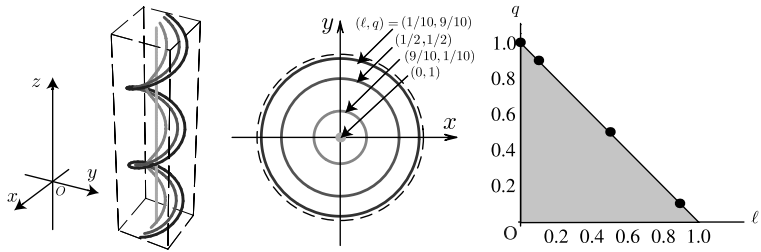}
\end{center}
\begin{minipage}{14cm}
\caption{Helical strings: 
$l + q =1  ~(q \neq 0)$. 
The three-dimensional snapshots are given in the left panel, and 
the projection of strings to the $x$-$y$ plane 
are given in the middle. 
The dashed circle in the middle figure represents the light cylinder 
$\rho=1/\Omega$. 
The parameters on the $l$-$q$ plane are plotted in the right panel. }
\label{fig:helical}
\end{minipage}
\end{figure}

We show here typical shapes of stationary rotating strings.
First we consider the case $ l +q = 1 ~(q \neq 0)$. 
The solutions are given by
\begin{equation}\begin{aligned}
	&\rho= \frac{\sqrt{ l }}{\Omega}, \quad
	&\varphi = \Omega (t + z).
\label{Helical_Wave}
\end{aligned}\end{equation}
In this case, a snapshot of string becomes a helix 
as shown in Fig. \ref{fig:helical}; 
we call these \lq helical strings\rq .

\bigskip

\begin{figure}[ht]
\centering
\bigskip
\includegraphics[width=14cm]{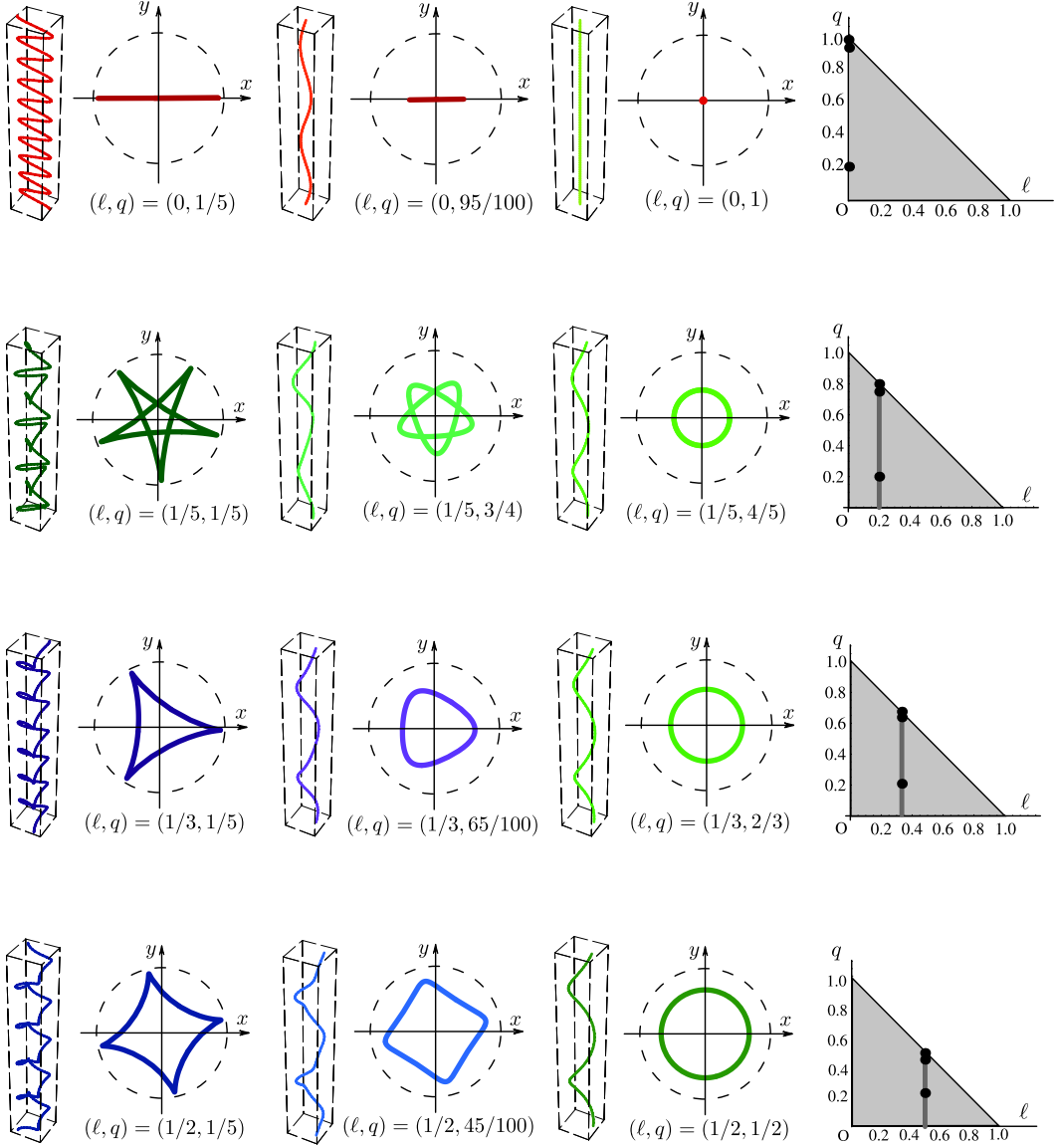}
\begin{minipage}{14cm}
\caption{
Three-dimensional snapshots and projections of string are shown in the 
case $l=0$, $1/5$, $1/3$, and $1/2$ as the same as Fig.\ref{fig:helical}. 
}
\label{fig:general}
\end{minipage}
 \end{figure}

Second we consider the case $ l  =0, \, q \neq 0 $. 
The solution can be described by
\begin{equation}
	x= \frac{\sqrt{1-q^2}}{\Omega} 
		\sin \left(\frac{\Omega z}{q}\right)\cos\Omega t, \quad
	y=\frac{\sqrt{1-q^2}}{\Omega} 
		\sin \left(\frac{\Omega z}{q}\right)\sin\Omega t,
\label{Plasol} 
\end{equation}
where 
$
	x:= \rho\cos\varphi, ~
 	y:= \rho \sin\varphi.
$
The strings, we call \lq planar\rq , are confined in a rotating plane.  
Snapshots of the planar strings are shown in the first row 
of Fig.\ref{fig:general}.

Third we consider the case 
$ l  + q \le 1 ~( l  \neq 0, q \neq 0) $.  
We show the shapes of strings in Fig.\ref{fig:general} 
for $ l  = 1/5,\, 1/3$, and $1/2$, respectively. 

If $l$ is a rational number, projection of the string on the $x$-$y$ plane 
becomes a closed curve. 
For $l=a/b (a,b~ \text{are relatively prime integer)}$, 
the closed curve consists 
of $N_l $ elements, where $N_l $ is defined by
\begin{equation}
	N_l = \frac{2b}{\mbox{GCD}[2b, (b-a)]}.
\end{equation}
Here, $\mbox{GCD}[a,b]$ denotes the greatest common divisor of $a,b$. 
The curve wraps around the center in the $x-y$ plane $M_l$ times 
until the curve returns to the starting point, where $M_l$ is 
given by
\begin{equation}
	M_l = \frac{1-l}{2}N_l ,
\label{Mdef}
\end{equation}
that is, 
\begin{equation}
	\bar\varphi(\sigma+N_l \sigmap)
		=\bar\varphi(\sigma)+2\pi M_l  ,
\label{period_phi}
\end{equation}
where $\sigmap:=\pi/\Omega$ is the periodicity of $\rho$ given by 
\eqref{SRSsol}.
The strings with rational $ l $ are periodic in $z$ with the period 
\begin{equation}
	\Zp = \pi N_l  q/\Omega. 
\label{string_period}
\end{equation}

\subsection{Energy, Momentum, and Angular Momentum} 

The string energy-momentum tensor $T^{\mu \nu}$ is given by\cite{VandS}  
\begin{align}
	\sqrt{-g} T^{\mu \nu}(x^\lambda)  
		&= -\mu \int d^2\zeta \Theta^{\mu \nu}(\zeta^c) 
				\delta^{(4)}\left(x^\lambda-x^\lambda(\zeta^c) \right), 
\label{EMTofCS} \\
	&\Theta^{\mu \nu}
		=\sqrt{-\gamma}\gamma^{ab}\partial_a x^\mu\partial_bx^\nu, 
\end{align}
where $x^\lambda(\zeta^c)$ is the solution of $\Sigma$. 
In the inertial reference system \eqref{MinMetCyl}, 
the explicit form of $\Theta^{\mu\nu}(\zeta^c)$, which depend only on 
$\sigma$, are shown in the Appendix. 

We define the string energy $E$, the angular momentum $J$, 
and the momentum along the rotation axis $P$. 
We consider infinitely long strings with periodic structure, i.e., $ l $ 
is assumed to be a rational number, 
then we define $E, J$ and $P$ for one period, $ z \sim  z+\Zp$ as 
\begin{align}
	E&:= \int_{\rho_\mathrm{min}}^{\rho_\mathrm{max}}d\rho\int_0^{2 \pi}d\varphi 
		\int_0^{\Zp} d z 
			\sqrt{-g}~T^{ t}_{\  \nu}(-\partial_{ t})^{\nu}
		= \mu \int_0^{N_l\sigmap}d\sigma\ 
			\Theta^{ t}_{~{ t}}(\sigma), 
\\
	J &:= \int_{\rho_\mathrm{min}}^{\rho_\mathrm{max}} d\rho 
			\int_0^{2 \pi} d\varphi \int_0^{\Zp} d z 
				\sqrt{-g}~ T^{ t}_{\  \nu}(\partial_{\varphi})^{\nu} 
		= -\mu \int_0^{N_l\sigmap} d\sigma \ 
			\Theta^{ t}_{~{\varphi}}(\sigma), 
\\
	P &:= \int_{\rho_\mathrm{min}}^{\rho_\mathrm{max}} d\rho\int_0^{2 \pi} d\varphi 
			\int_0^{\Zp} d z 
			\sqrt{-g}~T^{ t}_{\ \nu} (\partial_{ z})^{ \nu}
		= -\mu\int_0^{N_l\sigmap}d\sigma\ 
				\Theta^{ t}_{~ z}(\sigma).
\end{align}
We calculate these quantities as 
\begin{align}
	E &= \frac{\pi \mu }{|\Omega|} N_l  (1- l ^2), 
\label{EDef} \\
	J &= \frac{\pi \mu }{2\Omega |\Omega|} N_l  (1- l ^2 -q^2), 
\label{JDef} \\
	P &= - \frac{\pi\mu}{\Omega} N_l l q .
\label{PDef}   
\end{align}
Here, we take care of the sign of $\Omega, l$ and $q$ 
in \eqref{EDef}-\eqref{PDef}. 
We can also define the averaged values of these quantities per unit length of $z$ as 
\begin{align}
	\braket{E} &:= E/\Zp 
		= \mu \frac{ 1- l ^2 }{|q|} , 
\label{Eave}\\
	\braket{J} &:= J/\Zp 
		=\frac{\mu}{\Omega} \frac{ 1- l ^2 -q^2 }{2|q|} , 
\label{hEJPDef} \\
	\braket{P} &:= P/\Zp 
		=-\mu  l  \mathrm{sign}(\Omega q).
\label{Pave}
\end{align}
These quantities are applicable also for the strings with irrational $ l $.

The effective line density $\tilde\mu$, 
and effective tension $\tilde {\cal T} $ 
for the stationary rotating strings are defined\cite{VandS} 
in the reference system where the averaged value of 
momentum $\braket{P}$ vanishes. We obtain these quantities explicitly as
\begin{equation}\begin{aligned}
	&\tilde\mu 
		=\frac{\mu}{2|q|}
				\left[{1- l ^2+q^2}
				+\sqrt{(1-q- l )(1-q+ l )(1+q- l )(1+q+ l )}\right],\\
  &\tilde {\cal T} 
		=\frac{\mu}{2|q|}
				\left[{1- l ^2+q^2}
				-\sqrt{(1-q- l )(1-q+ l )(1+q- l )(1+q+ l )}\right].
\label{effectiveEOS}
\end{aligned}\end{equation}
In general, it holds that $\tilde \mu\tilde {\cal T}= \mu^2$
and $\tilde \mu \geq \tilde {\cal T}$. 
In the case of helical strings, there exists no inertial reference system 
such that $\braket{P}$ vanishes because a single wave moves with the velocity 
of light along the rotation axis.

\section{Gravitational perturbations} 
\subsection{Mode decomposition}

We consider metric perturbations $h_{\mu \nu}$ produced by a stationary 
rotating string in the Minkowski spacetime with the metric $\eta_{\mu \nu}$. 
We solve the linearized Einstein equations 
\begin{equation}
	\square \psi_{\mu \nu} 
		= -16\pi T_{\mu \nu},
\label{LEin} 
\end{equation}
where $T_{\mu \nu}$ are given by \eqref{EMTofCS}, and 
$\psi_{\mu \nu}$ is defined by
\begin{equation} 
	\psi_{\mu \nu} 
		= h_{\mu \nu} - \frac{1}{2} \eta_{\mu \nu} h_\alpha^\alpha. 
\end{equation}
We have used the Lorenz gauge condition 
$\partial^\mu \psi_{\mu \nu} =0$ in \eqref{LEin}.

We assume, here and henceforth, the parameter $l$ to be a rational number.
In this case, the stationary rotating string 
solutions \eqref{SRSsol} have periodicity in $z$ 
with the period $\Zp$ given by \eqref{string_period}. 
Then, $T_{\mu \nu}$ in \eqref{EMTofCS} have the following periodicities: 
\begin{align}
	T_{\mu \nu}(t,\rho, \varphi, z) 
		&= T_{\mu \nu} (t+{2\pi}/{\Omega} ,\rho, \varphi, z) , 
\\
	T_{\mu \nu}(t,\rho,\varphi,z) &= T_{\mu\nu}(t,\rho,\varphi+2\pi, z) , 
\\
	T_{\mu \nu}(t,\rho, \varphi, z) &= T_{\mu \nu}(t ,\rho, \varphi, z + \Zp) . 
\label{TPeriod} 
\end{align}
Thus, 
we can expand $T_{\mu \nu}$ in a Fourier series as 
\begin{align}
	T_{\mu \nu}(t,\rho, \varphi, z) 
		&= 
	\sum_{n=-\infty}^{\infty} \sum_{m=-\infty}^{\infty} 
	\sum_{s=-\infty}^{\infty} 
		e^{-i \omega_n t} \,e^{i m \varphi} \,e^{i k_s z} 
	 \tilde{T}_{\mu \nu}^{(n,m,s)} (\rho), 
\label{FExpT}
\end{align}
where
\begin{equation}
	\omega_n 
		:= \Omega  n, \quad
	k_s :=\frac{2\pi}{\Zp}s = \frac{2}{N_l} \frac{\Omega}{q} s, 
\end{equation}
and $n,m,s $ are integers. 

By using \eqref{EMTofCS}, 
we obtain the Fourier coefficients as
\begin{align}
	\tilde{T}_{\mu \nu}^{(n,m,s)} (\rho) 
		&= \frac{\Omega}{(2\pi)^2 \Zp}
			\int^{{2\pi}/{\Omega}}_{0} dt \int^{2 \pi}_0 
		d\varphi \int^{\Zp}_{0} dz 
		e^{i \omega_n t} \,e^{-i m \varphi} \,e^{-i k_s z} 
		T_{\mu \nu} (t,\rho , \varphi ,z) 
\\
	&= - \frac{ \mu \delta_{nm} }{2 \pi \Zp} 
		\int^{N_l\sigma_p}_{0} d\sigma 
 		e^{-i \left( k_s q \sigma + m \bar{\varphi} (\sigma) \right)}\, 
		\frac{1}{\rho} \Theta_{\mu\nu}(\sigma) 
		\delta (\rho - \rho_\mathrm{st}(\sigma)) ,  
\label{Ttilde2}
\end{align}
where $\rho_\mathrm{st}(\sigma)$ is the string 
solution given by \eqref{SRSsol}.
Because of $\delta_{nm}$ in \eqref{Ttilde2}, 
nonvanishing coefficients are specified by $(n, m)=(n, s)$, then we 
introduce $\tilde{T}_{\mu \nu}^{(n,s)}:= \tilde{T}_{\mu \nu}^{(n,n,s)}$.

We can also expand the metric perturbations $\psi_{\mu \nu}$ related to 
\eqref{FExpT} in a Fourier series as
\begin{equation}
	\psi_{\mu \nu} (t,\rho, \varphi, z) 
		= \sum_{n=-\infty}^{\infty}\sum_{s=-\infty}^{\infty} 
		e^{-i \omega_n t} \,e^{i n \varphi} \,e^{i k_s z} 
		\tilde{\psi}_{\mu \nu}^{(n,s)} (\rho) . 
\label{FExph} 
\end{equation}
Using \eqref{FExpT} and \eqref{FExph}, 
we can reduce \eqref{LEin} to a set of the ordinary differential equations 
with respect to $\rho$ for each Fourier mode labeled by $(n,s)$.

Ten components of linearized Einstein equations \eqref{LEin} are 
classified into three types: scalar type ($m=n$), vector type ($m=n \pm 1$), 
and tensor type ($m= n \pm 2$). 
Equations for these types have the following form: 
\begin{align}
\mbox{Scalar type}:&\quad 
	{\cal L}_{n}^{(n,s)} \,\tilde{\psi}_\mathrm{S}^{(n,s)} (\rho) 
		+16 \,\pi \,\tilde{T}_\mathrm{S}^{(n,s)} (\rho) =0 , 
\label{LhScalar} 
\\
\mbox{Vector type}:&\quad  
	{\cal L}_{n \pm 1}^{(n,s)} \,\tilde{\psi}_{\mathrm{V}\pm}^{(n,s)} (\rho) 
		+16 \,\pi \,\tilde{T}_{\mathrm{V}\pm}^{(n,s)} (\rho) =0 , 
\label{LhVector} 
\\
\mbox{Tensor type}:&\quad  
	{\cal L}_{n \pm 2}^{(n,s)} \,\tilde{\psi}_{\mathrm{T}\pm}^{(n,s)} (\rho) 
		+16 \,\pi \,\tilde{T}_{\mathrm{T}\pm}^{(n,s)} (\rho) =0,
\label{LhTensor} 
\end{align}
where the differential operator ${\cal L}_m^{(n,s)}$ 
with respect to $\rho$ is defined by 
\begin{equation}
	{\cal L}_m^{(n,s)} = 
	\frac{1}{\rho} \frac{d}{d\rho} \left( \rho \,\frac{d}{d \rho} \right) + 
	\left( \kappa_{ns}^2 - \frac{m^2}{{\rho}^2} \right)
\label{opL}
\end{equation}
with
\begin{equation}
	 \kappa_{ns}^2 := \omega_n^2 - k_s^2.
\label{kappa} 
\end{equation}
The members of 
$\left(\tilde{\psi}_\mathrm{S}^{(n,s)}, 
\tilde{\psi}_{\mathrm{V}\pm}^{(n,s)}, 
\tilde{\psi}_{\mathrm{T}\pm}^{(n,s)}\right)$ 
and 
$\left(\tilde{T}_\mathrm{S}^{(n,s)}, 
\tilde{T}_{\mathrm{V}\pm}^{(n,s)}, 
\tilde{T}_{\mathrm{T}\pm}^{(n,s)}\right)$
are defined by
\begin{equation}\begin{aligned}
	\tilde{\psi}_\mathrm{S}^{(n,s)} 
		&=	\left\{
		\tilde{\psi}_{tt}^{(n,s)}, 
		\tilde{\psi}_{zz}^{(n,s)}, 
		\tilde{\psi}_{tz}^{(n,s)}, 
		\left( \tilde{\psi}_{\rho \rho}^{(n,s)}  
				+ \tilde{\psi}_{\varphi \varphi}^{(n,s)}  /\rho^2 \right) 
	\right\},
\\
	\tilde{\psi}_{\mathrm{V}\pm}^{(n,s)}
		&=	\left\{
		\left(\tilde{\psi}_{t\rho}^{(n,s)} 
			\pm i\tilde{\psi}_{t\varphi}^{(n,s)}/\rho\right), 
		\left( \tilde{\psi}_{\rho z}^{(n,s)} 
			\pm i \tilde{\psi}_{\varphi z}^{(n,s)}/\rho \right) 
		\right\},
\\
	\tilde{\psi}_{\mathrm{T}\pm}^{(n,s)}
		&=	\left\{
		\left( \tilde{\psi}_{\rho \rho}^{(n,s)} 
			- \tilde{\psi}_{\varphi \varphi}^{(n,s)}/\rho^2 
			\pm 2 \, i \, \tilde{\psi}_{\rho \varphi}^{(n,s)}/\rho 
		\right) 
	\right\},
\end{aligned}\end{equation}
and
\begin{equation}\begin{aligned}
	\tilde{T}_\mathrm{S}^{(n,s)} 
		&= \left\{
		\tilde{T}_{tt}^{(n,s)} ,
		\tilde{T}_{zz}^{(n,s)} ,
		\tilde{T}_{tz}^{(n,s)} ,
		\left( \tilde{T}_{\rho \rho}^{(n,s)}  
				+ \tilde{T}_{\varphi \varphi}^{(n,s)}  /\rho^2 \right) 
	\right\},
\\
	\tilde{T}_{\mathrm{V}\pm}^{(n,s)} 
		&=\left\{
		\left( \tilde{T}_{t \rho}^{(n,s)}
			\pm i \, \tilde{T}_{t \varphi}^{(n,s)}/\rho \right), 
		\left( \tilde{T}_{\rho z}^{(n,s)} 
			\pm i \, \tilde{T}_{\varphi z}^{(n,s)}/\rho \right)
	\right\},
\\
	\tilde{T}_{\mathrm{T}\pm}^{(n,s)}
		&= \left\{
		\left( \tilde{T}_{\rho \rho}^{(n,s)} 
			- \tilde{T}_{\varphi \varphi}^{(n,s)}/\rho^2 
			\pm 2 \, i \, \tilde{T}_{\rho \varphi}^{(n,s)}/\rho 
		\right)
	\right\}, 
\label{T_type}
\end{aligned}\end{equation}
respectively.

At the infinity, because $m^2/\rho^2 \to 0$, \eqref{kappa} means 
the dispersion relation of the gravitational 
waves, where $\kappa_{ns}$ and $k_s$ can be regarded as 
the radial and the $z$ axis components of the wave vector, respectively.

\subsection{Green's function method} 

All of Eqs. \eqref{LhScalar}-\eqref{LhTensor} have the same form of
\begin{equation}
	{\cal L}_m^{(n,s)} \,\tilde{\psi}^{(n,s)} (\rho) 
			+16\pi \tilde{T}^{(n,s)} (\rho) =0, 
\label{Lh}
\end{equation}
where the indices $\mathrm{S, V\pm, T\pm}$ are suppressed. 
The ordinary differential equations \eqref{Lh} 
of the Sturm-Liouville type are 
formally solvable by using Green's function method. 
~(See~\cite{GeorgeArfken}, for example.)

Introducing Green's function $G_m^{ns}(\rho ,\rho')$ 
which satisfies 
\begin{equation}
	{\cal L}_m^{(n,s)} \,G_m^{ns} (\rho , \rho') = 
		- \frac{1}{\rho} \delta (\rho - \rho' ) \,, 
\label{Green1} 
\end{equation} 
we can express the solutions $\tilde{\psi}^{(n,s)}$ 
of \eqref{Lh} as
\begin{equation}
	\tilde{\psi}^{(n,s)} (\rho) 
		= \int_0^\infty 
	d\rho' G_{m}^{ns}(\rho, \rho') 16 \pi  \rho' \tilde{T}^{(n,s)}(\rho'). 
\label{LhSol} 
\end{equation}
Using \eqref{Ttilde2} for the scalar, vector, and tensor types of 
 $\tilde{T}^{(n,s)}$, we can write $\tilde{\psi}^{(n,s)}$ as
\begin{equation}
	\tilde{\psi}^{(n,s)}(\rho) 
		= - \frac{8 \mu}{q N_l \sigmap} 
	\int_0^{N_l \sigmap} 
	d\sigma \, G_{m}^{ns} (\rho , \rho_\mathrm{st}(\sigma)) \, \Theta(\sigma) 
	\exp \left( -i k_s q \sigma -i n \bar{\varphi}(\sigma) \right) , 
\label{LhSol2} 
\end{equation}
where $\Theta:=\{\Theta_\mathrm{S},\Theta_{\mathrm{V}\pm}, 
\Theta_{\mathrm{T}\pm}\}$ 
in the right hand side takes the same combination of 
$\Theta_{\mu\nu}$ as \eqref{T_type}. 
The coefficients $\tilde{\psi}^{(n,s)}$ should satisfy 
\begin{equation}
	\tilde{\psi}^{(-n,-s)}(\rho) 
	= \left(\tilde{\psi}^{(n,s)}(\rho)\right)^* ,
\label{reality}
\end{equation}
so that the metric perturbations $h_{\mu \nu}$ are real, 
where $*$ means the complex conjugate.

\subsection{Nonvanishing $(n,s)$ modes}

For the stationary rotating strings with rational $l$, 
the product $G^{ns}_m(\rho, \rho_\mathrm{st}(\sigma))\Theta(\sigma)$ 
in \eqref{LhSol2} is periodic in $\sigma$ 
with the period $\sigmap$ as 
\begin{equation}
	G^{ns}_m(\rho, \rho_\mathrm{st}(\sigma+\sigmap))\Theta(\sigma+\sigmap)
	=G^{ns}_m(\rho, \rho_\mathrm{st}(\sigma))\Theta(\sigma),
\end{equation}
because of the periodicity of $\rho_\mathrm{st}(\sigma)$ in \eqref{SRSsol}.
At the same time, from \eqref{period_phi} the exponential factor 
in \eqref{LhSol2} varies as 
\begin{align}
	\exp &\left( -i k_s q (\sigma+N_l\sigmap) 
		-i n \bar{\varphi}(\sigma+N_l\sigmap) \right)
\\
	&=\exp \left( -i k_s q \sigma -i n \bar{\varphi}(\sigma) \right) 
	\exp \left( -2\pi i 
		\left(s + n M_l \right) \right).
\end{align}
Here, we introduce a function $\Phi(\sigma)$ by
\begin{equation}
	\Phi(\sigma) 
		= \left(k_s q \sigma + n\bar\varphi(\sigma)\right)/L_{ns},
\end{equation}
where
\begin{equation}
	L_{ns}:=s + n M_l 
\end{equation}
is an integer specified by mode indices $n$ and $s$ 
for a stationary rotating string. 
The function $\Phi(\sigma)$ is monotonic in $\sigma$ and varies as
\begin{equation}
	\Phi(\sigma+N_l\sigmap) 
	=\Phi(\sigma) + 2\pi. 
\label{period_Phi}
\end{equation}
Then, Eq. \eqref{LhSol2} leads to
\begin{align}
	\tilde{\psi}^{(n,s)}(\rho) 
		&\propto 
	\int_0^{N_l \sigmap} 
	d\sigma \, G_{m}^{ns} (\rho , \rho_\mathrm{st}(\sigma)) \, \Theta(\sigma) 
	\exp \left( -i L_{ns}\Phi(\sigma) \right) 
\cr
		&=  
	\int_0^{2\pi} d\Phi
	\frac{d\sigma}{d\Phi} 
	 G_{m}^{ns}(\rho, \rho_\mathrm{st}(\sigma(\Phi)))\Theta(\sigma(\Phi)) 
	\exp\left( -i L_{ns}\Phi\right),
\label{expression_psi}
\end{align}
where we have changed the integration variable $\sigma$ by $\Phi$. 
Since $d\bar\varphi/d\sigma$ is periodic with the period $\sigmap$, 
$d\Phi/d\sigma$ is also periodic in $\sigma$ with the same period. 
Therefore, we can see that 
$d\sigma/d\Phi$, 
$\rho_\mathrm{st}(\sigma(\Phi))$, and $\Theta(\sigma(\Phi))$ have 
a periodicity in $\Phi$ with the period $2\pi/N_l$, 
and then, we can obtain a Fourier series 
\begin{equation}
\frac{d\sigma}{d\Phi}
	G_m^{ns} (\rho,\rho_\mathrm{st}(\sigma(\Phi)))\Theta(\sigma(\Phi))	
	=	\sum_j a_j \exp\left( i ~j N_l\Phi\right) , 
\end{equation}
where $a_j$ are Fourier coefficients labeled by an integer $j$. 
Inserting this into \eqref{expression_psi} we have
\begin{equation}
	\tilde{\psi}^{(n,s)}(\rho) 
		\propto 
	\int_0^{2\pi} d\Phi
	\sum_j a_j
	\exp\left( i (j N_l-L_{ns})\Phi\right). 
\end{equation}
Therefore, 
for the combination $(n,s)$ of nonvanishing $\tilde{\psi}^{(n,s)}(\rho)$, 
there should exist an integer $j$ which satisfies 
\begin{equation}
	L_{n s}= s + n M_l 	=j N_l .
\label{OthCon2} 
\end{equation}

Especially, in the case of helical strings, $l+q=1, q \neq 0$, 
because 
$G_{m}^{ns} (\rho , \rho_\mathrm{st}(\sigma))  \Theta(\rho_\mathrm{st}(\sigma)) $ 
in \eqref{LhSol2} is constant with respect to $\sigma$, 
 the nonvanishing $(n,s)$-modes is specified by the condition 
\begin{equation}
 	L_{ns}= s + n M_l =0 . 
\label{ConHel} 
\end{equation}

\subsection{Explicit forms of Green's functions} 

We obtain the explicit form of Green's functions $G_{m}^{ns}$, here. 
We consider three cases with respect to the sign of $\kappa_{ns}^2$ 
defined by \eqref{kappa}. 

First, we consider the case $\kappa_{ns}^2 < 0$. 
If we require the regularity both at the center and at the infinity, 
the operator 
\eqref{opL} with negative $\kappa_{ns}^2$ allows 
damping solutions to \eqref{Lh} with the length scale 
$|\kappa_{ns}^{-1}|$. 
Green's functions in this case have the form
\begin{equation}
	{G^{ns}_{m}} (\rho ,\rho') = 
		I_m (|\kappa_{ns}|\rho) K_m (|\kappa_{ns}|\rho') \theta(\rho'-\rho) 
		+ K_m (|\kappa_{ns}|\rho)I_m(|\kappa_{ns}|\rho')\theta(\rho-\rho'), 
\label{GMB}
\end{equation}
where the functions  $\theta(x)$ is the Heviside step function, and 
$I_m (x) $ and $ K_m (x) $ are the modified Bessel functions, 
\begin{align}
	I_m (x) = i^{-m} J_m (ix), \quad
	K_m (x) = (\pi /2) \, i^{m+1} H_m^{(1)} (i x) ,
\end{align}
and $J_m$ and $H_m^{(1)}$ are the Bessel function 
and the Hankel function of the first kind, respectively. 

Next, in the case $\kappa_{ns}^2=0$, 
because the scale vanishes in the operator 
\eqref{opL}, the solutions to \eqref{Lh} have long tails. 
Green's functions are 
\begin{align}
	{G^{ns}_{m}} (\rho ,\rho') 
		&= \begin{cases}
		-\left\{ \ln (\rho' / \rho_0) \, \theta (\rho' - \rho ) 
		+\ln (\rho / \rho_0) \, \theta (\rho - \rho' )
 		\right\} &\mbox{for $m=0$} , 
\\
	 \frac{1}{2 |m|} 
		\left\{ \left( \rho/\rho' \right)^{|m|} \theta(\rho'-\rho ) 
		+ \left( \rho'/\rho \right)^{|m|} \theta (\rho-\rho' )
 	\right\} &\mbox{for $m\neq 0$} . 
\end{cases}
\label{GPL}
\end{align}
In the case of $m=0$, we have introduced 
a constant $\rho_0$ as the boundary instead of the infinity such that 
$G^{ns}_{0} \to 0$ in the limit $\rho \to \rho_0$.

Finally, in the case $\kappa_{ns}^2 > 0$, 
the operator \eqref{opL} allows wave solutions to \eqref{Lh}. 
The scale $\kappa_{ns}^{-1}$ gives the wave length of the solutions. 
Green's functions take the form of 
\begin{equation}
	{G^{ns}_{m}} (\rho ,\rho') 
		= \frac{\pi }{2} \,i \left\{
		J_m (\kappa_{ns} \rho) H_m (\kappa_{ns} \rho')\theta(\rho'-\rho) 
		+H_m (\kappa_{ns}\rho)J_m (\kappa_{ns} \rho')\theta(\rho-\rho')
	\right\}.
\label{GB}
\end{equation}
Here, $H_m$ are defined by
\begin{align}
 H_m (x) 
	& = 
		\left\{ \begin{array}{rl}
  		H_{m}^{(1)} (x) & \mbox{for $\omega_n > 0$} ,\\
 		- H_m^{(2)} (x) & \mbox{for $\omega_n < 0$} , 
       \end{array} 
\right.  
\end{align}
where $H_m^{(1)}$ and $H_m^{(2)}$ denote
the Hankel functions of first and second kind, respectively. 
This definition guarantees that the solutions describe the out-going waves 
at the infinity in any case of $\omega_n$.

\subsection{Potential mode and wave modes}

\begin{figure}[ht]
\centering
\includegraphics[width=6cm]{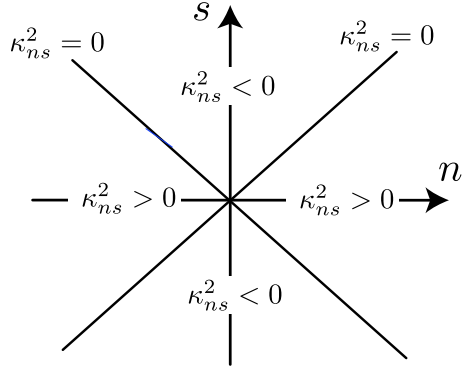} 
\caption{Three cases of $\kappa_{ns}^2$ in the $n$-$s$ plane.}
 \label{nsFIG}
\end{figure}

In the previous subsection, Green's functions 
are constructed in three different cases: 
$\kappa_{ns}^2 <0$, $\kappa_{ns}^2 =0$, and $\kappa_{ns}^2 >0$, respectively. 
These three cases correspond to the regions on the $n$-$s$ plane as
\begin{align}
	|s| &> \left|\frac{q N_l}{2} n\right|
		\quad \mbox{for} \quad 
	\kappa_{ns}^2 <0 , 
\label{negative_kappa}\\
	|s| &= \left|\frac{q N_l}{2} n\right|
		\quad \mbox{for} \quad 
	\kappa_{ns}^2 =0 , 
\label{zero_kappa}\\
	|s| &< \left|\frac{q N_l}{2} n\right|
		\quad \mbox{for} \quad 
	\kappa_{ns}^2 >0 , 
\label{positive_kappa}
\end{align}
which are shown in Fig.\ref{nsFIG}. 

The two lines which denote $\kappa_{ns}^2=0$ in the $n$-$s$ plane 
are given by
\begin{equation}
	s = \pm \frac{q N_l}{2} n. 
\end{equation}
The inclinations of the lines, which depend on $l$ and $q$,  
have the maximum absolute value $M_l$ when $q=1-l$ for given $l$.

\begin{figure}[!ht]
\centering
\includegraphics[width=15cm]{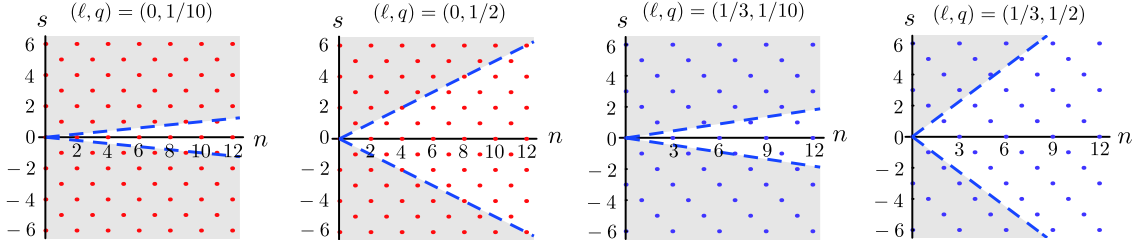} 
\begin{minipage}{14cm}
\caption{
The pairs $(n,s)$ of nonvanishing modes specified by \eqref{OthCon2} 
are shown by dots for $l=0$ and $l=1/3$ cases, as examples. 
The dots in the shadowed regions are short-range modes, 
while the dots in the unshaded region represent the gravitational wave modes. 
The dots on the border, thick broken lines, are traveling wave modes, and 
$(n,s)=(0,0)$ is the gravitational potential mode. 
 \label{ms}
 }
\end{minipage}
\end{figure}

Here, we divide the metric perturbation into four parts, 
namely short range force modes $h_{\mu \nu}^\mathrm{Short}$, 
stationary potential mode $h_{\mu \nu}^\mathrm{Pot}$, 
traveling wave modes $h_{\mu \nu}^\mathrm{TW}$, 
and gravitational wave modes $h_{\mu \nu}^\mathrm{GW}$ as
\begin{equation}
	h_{\mu \nu} 
		= h_{\mu \nu}^\mathrm{Short} + h_{\mu \nu}^\mathrm{Pot} 
 		+ h_{\mu \nu}^\mathrm{TW} + h_{\mu \nu}^\mathrm{GW},
\end{equation}
where 
\begin{align}
	&h_{\mu \nu}^\mathrm{Short} (t,\rho, \varphi, z) 
		:= \sum_{\substack{n>0\\ |s|>(q N_l/2)n}} 
	\left[ \exp \left( -i \omega_n t + i n \varphi + i k_s z \right)
		\tilde{h}_{\mu \nu}^{(n,s)} (\rho) + \mbox{(c.c.)}\right], 
\label{short} \\
	&h_{\mu \nu}^\mathrm{Pot} (t,\rho, \varphi, z) 
		:= \tilde{h}_{\mu \nu}^{(0,0)} (\rho) ,
\label{SP} \\
	&h_{\mu \nu}^\mathrm{TW} (t,\rho, \varphi, z) 
		:= \sum_{\substack{n>0\\ |s|=(q N_l/2)n}} 
	\left[ \exp \left( -i \omega_n t + i n \varphi + i k_s z \right)
		\tilde{h}_{\mu \nu}^{(n,s)} (\rho) +\mbox{(c.c.)} \right],
\label{DP} \\
	&h_{\mu \nu}^\mathrm{GW} (t,\rho, \varphi, z) 
		:= \sum_{\substack{n>0\\ |s|<(q N_l/2)n}} 
	\left[ \exp \left( -i \omega_n t + i n \varphi + i k_s z \right)
		 \tilde{h}_{\mu \nu}^{(n,s)} (\rho) +\mbox{(c.c.)} \right], 
\label{FExphGW} 
\\
&\hspace{5cm}(\mbox{(c.c.) denotes complex conjugate}). \nonumber
\end{align}
The summations in \eqref{short}, \eqref{DP} and \eqref{FExphGW} are taken over 
pairs $(n, s)$ which satisfy the condition \eqref{OthCon2} or \eqref{ConHel}. 
These $(n, s)$ are shown in Fig.\ref{ms} as dots in the $n$-$s$ plane.

The modes in $\kappa_{ns}^2<0 ~ (|s|>(q N_l/2)n)$, given by 
Green's function \eqref{GMB}, 
describe the gravitational field 
in a short range around the string, 
and exponentially decrease in the region $\rho \gg {1/|\Omega|}$; 
we name these \lq{\sl short-range modes}\rq. 
Since a distant observer hardly access the short-range modes, 
we do not discuss these further. 

The mode of $(n,s)=(0,0)$ is clearly time independent. 
The metric components of this mode represents the Newtonian potential, 
the angular deficit, and the effects of frame-dragging.   
The modes in $\kappa_{ns}^2=0$ describe waves propagating along 
the $z$ axis, i.e., along the rotating string. 
These waves are named \lq{\sl traveling waves}\rq\ following 
Ref.\cite{Vachaspati}. 
The modes in $\kappa_{ns}^2>0$ are gravitational waves propagating 
toward distant observers from the string. 
The facts noted above will be in successive sections. 

\section{Potential mode}

The mode $(n,s)=(0,0)$ describes time-independent long range potential. 
The components $\tilde{\psi}_{\mu \nu}^{(0,0)}$ given 
by \eqref{LhSol2} have the following form:
\begin{equation}\begin{aligned}
  &\tilde{\psi}^{(0,0)}_\mathrm{S} (\rho) 
	= -\frac{8\mu }{q N_l \sigma_\mathrm{p}}
		\int_{0}^{N_l \sigma_\mathrm{p}} d\sigma 
		G_{0}^{00}(\rho, \rho_\mathrm{st}(\sigma)) \Theta_\mathrm{S}(\sigma) , 
\\
  &\tilde{\psi}^{(0,0)}_\mathrm{V\pm} (\rho) 
	= -\frac{8\mu }{q N_l \sigma_\mathrm{p}}
		\int_{0}^{N_l \sigma_\mathrm{p}} d\sigma 
		G_{\pm1}^{00}(\rho, \rho_\mathrm{st}(\sigma))\Theta_\mathrm{V\pm}(\sigma),
\\
  &\tilde{\psi}^{(0,0)}_\mathrm{T\pm} (\rho) 	
	= -\frac{8\mu }{q N_l \sigma_\mathrm{p}}
		\int_{0}^{N_l \sigma_\mathrm{p}} d\sigma 
		G_{\pm2}^{00}(\rho, \rho_\mathrm{st}(\sigma))\Theta_\mathrm{T\pm}(\sigma), 
\label{PLLogSol} 
\end{aligned}
\end{equation}
where Green's functions are given by \eqref{GPL}. 
After some calculations, explicit forms of
$
	h^\mathrm{Pot}_{\mu\nu} (\rho)=h^{(0,0)}_{\mu\nu} (\rho)
$
in the far region are given as  
\begin{align}
	h^\mathrm{Pot}_{tt} 
		&= h^\mathrm{Pot}_{zz}  
		= -\frac{4\mu}{q}(1-l^2-q^2)\ln\left( \frac{\rho}{\rho_0}\right),
 \\
	h^\mathrm{Pot}_{tz}  
		&= -8\mu l 
		\ln \left( \frac{\rho}{\rho_0} \right), 
\label{dragz}
\\
	h^\mathrm{Pot}_{\rho \rho} 
 &= -\frac{4\mu}{q}(1-l^2 +q^2) \ln \left( \frac{\rho}{\rho_0} \right) 
	  - \frac{\mu}{q}\left((1-l^2-q^2)^2-4l^2q^2\right)
		\frac{1}{(\Omega\rho)^2} , 
\\
	\frac{h^\mathrm{Pot}_{\varphi \varphi} }{ \rho^2} 
 &= -\frac{4\mu}{q}(1-l^2 +q^2) \ln \left( \frac{\rho}{\rho_0} \right) 
	  + \frac{\mu}{q}\left((1-l^2-q^2)^2-4l^2q^2\right)
		\frac{1}{(\Omega\rho)^2} , 
\\ 
	\frac{h^\mathrm{Pot}_{t\varphi} }{ \rho} 
		&= -\frac{2\mu}{q}(1-l^2 -q^2) \frac{1}{\Omega  \rho },  \\
	\frac{h^\mathrm{Pot}_{z \varphi}}{\rho }
		&=  -4\mu l  \frac{1}{\Omega\rho},  \\
	h^\mathrm{Pot}_{t \rho} 
		&= h^\mathrm{Pot}_{z\rho} 
		= h^\mathrm{Pot}_{\rho \varphi}/\rho = 0. 
\label{PotSol}
\end{align}
Although we assume $l $ to be a rational number, 
the expressions of $h_{\mu \nu}^\mathrm{Pot}$ given above 
are also valid for irrational $l$.

It is found that $h^\mathrm{Pot}_{t\varphi}$ denotes the 
azimuthal frame-dragging 
caused by the angular momentum of the string, 
and $h^\mathrm{Pot}_{tz}$ does 
dragging along the $z$ axis caused by the 
linear momentum along the rotation axis of the string. 
In the case of planar strings, $l = 0$, we see that $\braket{P}=0$ from 
\eqref{Pave} and that there is no dragging along 
the $z$ axis from \eqref{dragz}. 
If we transform the inertial reference frame 
$(t,\rho,\varphi,z)\to (\tilde{t}, \tilde{\rho}, \tilde{\varphi}, \tilde{z} )$ 
by the Lorentz boost such that $\braket{P}=0$ as shown in Ref.\cite{RRS},  
the dragging along $z$ axis disappears. 
In this frame, the logarithmic terms of 
$h_{\mu\nu}^\mathrm{Pot}$ give the metric in the form: 
\begin{align}
	ds^2 
	&= -\left(1+4(\tilde{\mu}-\tilde{\cal T})\ln(\tilde\rho/\rho_0)\right)
		d\tilde t^2
		+ \left(1-4(\tilde{\mu}-\tilde{\cal T})\ln(\tilde\rho/\rho_0)\right)
				d\tilde z^2 
\cr & \hspace{2cm}
		+\left(1-4(\tilde{\mu}+\tilde{\cal T}) \ln(\tilde\rho /\rho_0)\right) 
			(d\tilde\rho^2 + \tilde\rho^2 d\tilde \varphi^2 ) .
\label{metric_pot} 
\end{align}
Using the coordinate transformation,  
\begin{equation}
	  r =\left(1+ 2(\tilde{\mu}+\tilde{\cal T})
		(1- \ln (\tilde\rho /\rho_0)\right)\tilde\rho, 
\quad
	\phi= 	\left(1- 2(\tilde{\mu}+\tilde{\cal T})\right)\tilde\varphi ,
\end{equation}
and ignoring ${\cal O}\left((\tilde\mu+\tilde{\cal T})^2\right)$ terms, 
the metric of the $\tilde t=const.$ and $\tilde z=const.$ surface becomes 
flat metric
\begin{equation}
	ds^2 =dr^2 + r^2 d\phi^2. 
\end{equation}
Since the range of $\phi$ is 
$0 \leq \phi<2\pi(1- 2(\tilde{\mu}+\tilde{\cal T}))$, 
the flat surface 
is the conical space with 
angular deficit $4 \pi (\tilde{\mu}+\tilde{\cal T})$~\cite{VandS}. 

Alternatively, using the coordinate
\begin{equation}
 	\bar r = \left(1+4\tilde{\cal T}
		(1-\ln(\tilde\rho/\rho_0)\right)\tilde\rho, 
\quad
	\bar{\phi} = (1-4\tilde{\cal T}) \tilde\varphi, 
\end{equation} 
we rewrite the metric \eqref{metric_pot} as 
\begin{equation}
	ds^2 = - \left(1 + 2 \Psi(\tilde\rho) \right) d\tilde t^2 
		+ \left(1 - 2 \Psi(\tilde\rho) \right) 
	\left( d\bar r^2 +\bar r^2 d\bar{\phi}^2 +d\tilde z^2 \right)
\end{equation}
where 
\begin{equation}\begin{aligned}
	\Psi(\tilde\rho) = 2(\tilde{\mu}-\tilde{\cal T}) \ln(\tilde\rho /\rho_0). 
\end{aligned}\end{equation}
This metric 
means that 
the stationary rotating string produces the Newtonian logarithmic 
potential $\Psi$ around it~\cite{VandS}. 

In general, the stationary rotating string in the frame of $\braket{P}=0$ 
yields the logarithmic potential, the angular deficit, 
and the azimuthal frame-dragging in $\varphi$. 
It should be noted, as an exceptional case, that the dragging along 
the rotation axis, the $z$ axis, 
can not be erased for the helical strings because there is no reference frame 
such that $\braket{P}=0$. 
In addition, the Newtonian potential vanishes, and the angular deficit, 
$8\pi\mu$, is the same value as the straight string.

\section{Gravitational Wave Modes} 

In this section, we consider the metric perturbations propagating  
away from a string to a distant observer, 
i.e., the gravitational wave modes ~$h_{\mu \nu}^\mathrm{GW}$   
given in \eqref{FExphGW}, 
where the summation is taken over all $(n,s)$ satisfying \eqref{OthCon2} 
and \eqref{positive_kappa}.
Fourier components of metric perturbations 
$\tilde{h}_{\mu \nu}^{(n,s)} (\rho)$, 
equivalently $\tilde{\psi}_{\mu \nu}^{(n,s)} (\rho)$, 
are given by \eqref{LhSol2} where Green's functions are \eqref{GB}.

First, 
we define the physical modes of polarization, 
plus-modes and cross-modes. 
Next, we show that the gravitational waves can be 
emitted to several discrete directions. 
Finally, we present waveforms of 
the gravitational waves emitted to the possible directions 
by using numerical calculations.

\subsection{Plus modes and cross modes}

Here, we fix the gauge freedom of propagating modes in the vacuum. 
We use the transverse traceless (TT) gauge conditions:
\begin{equation}
	h^{TT}_{t\mu}=0,\quad \partial^i h^{TT}_{ij}=0,\quad {h^{TT}}^i_{~i}=0.  
\label{TTgauge}
\end{equation}
The metric perturbations satisfying TT-conditions, $h^\mathrm{TT}_{ij}$, 
are invariant under gauge transformations. 
Using the fact that the Riemann tensor, which is gauge invariant, 
is expressed by $h^\mathrm{TT}_{ij}$ in the linear order, 
we can obtain the TT-modes by integration of 
\begin{equation}
	\partial_t^2 h_{ij}^\mathrm{TT} = -2 R_{i t j t} 
		=-\left( 
		\partial_t \partial_j h_{it} 
		+ \partial_i \partial_t h_{tj} 
			-\partial_i \partial_j h_{t t } 
		-\partial_t \partial_t h_{ij}
		 \right) ,
\end{equation}
where $h_{\mu\nu}$ in the right hand side are solutions of the wave 
equation. 
(See the Sec. 35.4 of \cite{Gravitation}.) 

In the cylindrical coordinate, $\tilde{h}_{ij }^{(n,s)\mathrm{TT}}$ 
can be obtained as 
\begin{align}
\tilde{h}_{\rho \rho }^{(n,s)\mathrm{TT}} 
	&=  \frac{1}{2} 
	\left( \tilde{\psi}_{tt}^{(n,s)} 
	+ \tilde{\psi}_{\rho \rho}^{(n,s)} 
	- \frac{\tilde{\psi}_{\varphi \varphi}^{(n,s)}}{\rho^2} 
	- \tilde{\psi}_{zz}^{(n,s)} \right) 
	- \frac{2 i}{\omega_n} \partial_\rho \tilde{\psi}_{t \rho}^{(n,s)} 
\cr 
	&\hspace{1cm}-\frac{1}{2 \omega_n^2} 
	\partial_\rho^2 \left( \tilde{\psi}_{tt}^{(n,s)} 
	+\tilde{\psi}_{\rho \rho}^{(n,s)} 
	+\frac{ \tilde{\psi}_{\varphi \varphi}^{(n,s)}}{\rho^2} 
	+\tilde{\psi}_{zz}^{(n,s)} \right) , 
\cr
\frac{\tilde{h}_{\rho \varphi }^{(n,s)\mathrm{TT}} }{\rho}
	&= \frac{\tilde{\psi}_{\rho \varphi }^{(n,s)}}{\rho} 
	- \frac{i}{\omega_n} \left( \partial_\rho - \frac{1}{\rho} \right) 
	\left( \frac{\tilde{\psi}_{t \varphi }^{(n,s)}}{\rho} \right) 
\cr 
	&\hspace{1cm}+ \frac{1}{\Omega \rho} \left\{ 
	\tilde{\psi}_{t \rho }^{(n,s)} - \frac{i}{2 \omega_n} 
	\left( \partial_\rho - \frac{1}{\rho} \right) 
	\left( \tilde{\psi}_{tt}^{(n,s)} 
	+\tilde{\psi}_{\rho \rho}^{(n,s)} 
	+\frac{ \tilde{\psi}_{\varphi \varphi}^{(n,s)}}{\rho^2} 
	+\tilde{\psi}_{zz}^{(n,s)} \right) 
	\right\} , 
\cr
\tilde{h}_{\rho z}^{(n,s)\mathrm{TT}} 
	&= \tilde{\psi}_{\rho z }^{(n,s)} 
	- \frac{i}{\omega_n} \partial_\rho \tilde{\psi}_{tz}^{(n,s)} 
	+\frac{k_s}{\omega_n} \left\{
	\tilde{\psi}_{t \rho }^{(n,s)}
	- \frac{i}{2 \omega_n} \partial_\rho 
	\left( \tilde{\psi}_{tt}^{(n,s)} 
	+\tilde{\psi}_{\rho \rho}^{(n,s)} 
	+\frac{ \tilde{\psi}_{\varphi \varphi}^{(n,s)}}{\rho^2} 
	+\tilde{\psi}_{zz}^{(n,s)} \right) 
	\right\} , 
\cr
\frac{\tilde{h}_{\varphi \varphi}^{(n,s)\mathrm{TT}} }{\rho^2}
	&=  \frac{1}{2} 
	\left( \tilde{\psi}_{tt}^{(n,s)} 
	- \tilde{\psi}_{\rho \rho}^{(n,s)} 
	+ \frac{\tilde{\psi}_{\varphi \varphi}^{(n,s)}}{\rho^2} 
	- \tilde{\psi}_{zz}^{(n,s)} \right) 
	+ \frac{2}{\Omega \rho} 
	\left\{ \frac{\tilde{\psi}_{t \varphi}^{(n,s)}}{\rho}  
	- \frac{i}{n} \tilde{\psi}_{t \rho }^{(n,s)}
	 \right\} 
\cr
	&\hspace{1cm}+ \frac{1}{2(\Omega \rho)^2} 
	\left( 1 -\frac{\rho}{n^2} \partial_\rho \right) 
	\left( \tilde{\psi}_{tt}^{(n,s)} 
	+\tilde{\psi}_{\rho \rho}^{(n,s)} 
	+\frac{ \tilde{\psi}_{\varphi \varphi}^{(n,s)}}{\rho^2} 
	+\tilde{\psi}_{zz}^{(n,s)} \right) , 
\cr
\frac{\tilde{h}_{\varphi z}^{(n,s)\mathrm{TT}} }{\rho}
	&= \frac{\tilde{\psi}_{\varphi z}^{(n,s)} }{\rho} 
	+\frac{k_s}{\omega_n} \left( \frac{\tilde{\psi}_{t \varphi}^{(n,s)}}{\rho} 
	\right) 
	+ \frac{1}{\Omega \rho} \left\{ 
	\tilde{\psi}_{tz}^{(n,s)} 
	+ \frac{k_s}{2 \omega_n} 
	\left( \tilde{\psi}_{tt}^{(n,s)} 
	+\tilde{\psi}_{\rho \rho}^{(n,s)} 
	+\frac{ \tilde{\psi}_{\varphi \varphi}^{(n,s)}}{\rho^2} 
	+\tilde{\psi}_{zz}^{(n,s)} \right)
	\right\} , 
\cr
\tilde{h}_{zz}^{(n,s)\mathrm{TT}} 
	&= \frac{1}{2} 
	\left( \tilde{\psi}_{tt}^{(n,s)} 
	-\tilde{\psi}_{\rho \rho}^{(n,s)} 
	-\frac{ \tilde{\psi}_{\varphi \varphi}^{(n,s)}}{\rho^2} 
	+\tilde{\psi}_{zz}^{(n,s)} \right) 
	+ \frac{2 k_s}{\omega_n} \tilde{\psi}_{tz}^{(n,s)} 
\cr
	&\hspace{1cm} +\frac{1}{2} \left( \frac{k_s}{\omega_n} \right)^2 
	\left( \tilde{\psi}_{tt}^{(n,s)} 
	+\tilde{\psi}_{\rho \rho}^{(n,s)} 
	+\frac{ \tilde{\psi}_{\varphi \varphi}^{(n,s)}}{\rho^2} 
	+\tilde{\psi}_{zz}^{(n,s)} \right) . 
\label{tildehTT}
\end{align}

In the large distance limit, 
the wave vector of a $(n,s)$-mode in the normalized orthogonal frame 
$(\hat t, \hat \rho, \hat\varphi, \hat z)$ is expressed as 
\begin{align}
	\hat k_{\hat\mu}^{(n,s)} = ( -\omega_n , \kappa_{ns} , 0 , k_s ) \,,
\end{align}
because the $\hat\varphi$-component of the wave vector becomes small 
as $1/\rho$ in the far region. 
Then, the $(n,s)$-mode 
propagates in the direction specified by the angle $\theta_{s/n}$ 
from the rotation axis which is defined by
\begin{equation}
	\cos \theta_{s/n} =\frac{k_s }{\omega_n} 
	=\frac{2}{N_l q}\frac{s}{n}. 
\label{theta} 
\end{equation} 
The direction $\theta_{s/n}=\pi/2$ is perpendicular to the $z$ axis, i.e., 
perpendicular to the string. 

Here, we introduce a new normal frame 
$( \hat t, \hat\eta, \hat\varphi, \hat\zeta)$ at the observer, 
such that the direction of the wave vector coincides with $\hat \eta$, i.e., 
\begin{equation}
	-\hat k^{(n,s)}_{\hat t} = \hat k^{(n,s)}_{\hat\eta} =\omega_n. 
\end{equation}
The new basis is defined for each $(n,s)$ explicitly as
\begin{align}
	&\hat\eta =\sin\theta_{s/n} \hat\rho + \cos\theta_{s/n} \hat{z}, 
\\
	&\hat\zeta=-\cos\theta_{s/n} \hat\rho+\sin\theta_{s/n} \hat{z}.
\end{align}
By the use of this frame the components of metric perturbations 
\eqref{tildehTT} are given by 
\begin{align}
	\tilde{h}_{\hat\eta \hat\eta}^{(n,s)\mathrm{TT}} 
	 &= 
		\left(\frac{\kappa_{ns}}{\omega_n}\right)^2 
			\tilde{h}_{\rho \rho }^{(n,s)\mathrm{TT}}
			+2 \left( \frac{k_{s} \kappa_{ns}}{\omega_n^2} \right) 
				\tilde{h}_{\rho z}^{(n,s)\mathrm{TT}} 
			+ \left( \frac{k_{s}}{\omega_n} \right)^2 
				\tilde{h}_{zz}^{(n,s)\mathrm{TT}}, 
\cr
	\tilde{h}_{\hat\zeta \hat\zeta}^{(n,s)\mathrm{TT}} 
		&= \left( \frac{k_{s}}{\omega_n} \right)^2 
			\tilde{h}_{\rho \rho }^{(n,s)\mathrm{TT}}
		-2 \left( \frac{k_{s} \kappa_{ns}}{\omega_n^2} \right) 
			\tilde{h}_{\rho z}^{(n,s)\mathrm{TT}} 
		+ \left( \frac{\kappa_{ns}}{\omega_n} \right)^2 
			\tilde{h}_{zz}^{(n,s)\mathrm{TT}} ,
\cr
	\tilde{h}_{\hat\varphi\hat\varphi}^{(n,s)\mathrm{TT}} 
		&= 
		\frac{\tilde{h}_{\varphi\varphi}^{(n,s)\mathrm{TT}}}{\rho^2}, 
\cr
	\tilde{h}_{\hat\eta \hat\zeta}^{(n,s)\mathrm{TT}} 
		&= 
		\left\{ 1- 2 \left( \frac{k_{s}}{\omega_n} \right)^2 \right\} 
			\tilde{h}_{\rho z }^{(n,s)\mathrm{TT}}
		- \left( \frac{k_{s} \kappa_{ns}}{\omega_n^2} \right) 
			\left( \tilde{h}_{\rho \rho}^{(n,s)\mathrm{TT}} 
		- \tilde{h}_{zz}^{(n,s)\mathrm{TT}} \right) ,
\cr
	\tilde{h}_{\hat\eta \hat\varphi}^{(n,s)\mathrm{TT}}
		&= 
		\left( \frac{\kappa_{ns}}{\omega_n} \right) 
		\frac{\tilde{h}_{\rho \varphi }^{(n,s)\mathrm{TT}}}{\rho} 
		+ \left( \frac{k_{s}}{\omega_n} \right) 
			\frac{\tilde{h}_{\varphi z}^{(n,s)\mathrm{TT}}}{\rho} ,
\cr
	\tilde{h}_{\hat\varphi \hat\zeta}^{(n,s)\mathrm{TT}}
		&= 
		- \left( \frac{k_{s}}{\omega_n} \right) 
			\frac{\tilde{h}_{\rho \varphi }^{(n,s)\mathrm{TT}}}{\rho} 
		+ \left( \frac{\kappa_{ns}}{\omega_n} \right) 
			\frac{\tilde{h}_{\varphi z}^{(n,s)\mathrm{TT}}}{\rho} \,. 
\label{tildehTT2}
\end{align}
It can be shown that $\tilde{h}_{\hat\eta \hat\eta }^{(n,s)\mathrm{TT}}$, 
$\tilde{h}_{\hat\eta \hat\varphi }^{(n,s)\mathrm{TT}}$, and 
$\tilde{h}_{\hat\eta \hat\zeta }^{(n,s)\mathrm{TT}}$ are vanishing 
by using the wave equation, i.e., \eqref{kappa}, and TT-gauge condition 
\eqref{TTgauge}. 
For convenience, we define the two modes of polarizations: 
the plus-mode $\tilde{h}_{+}^{(n,s)}$, and 
the cross-mode $\tilde{h}_{\times}^{(n,s)}$, as 
\begin{align}
	\tilde{h}_{+}^{(n,s)} (\rho ) 
		&=\tilde{h}_{\hat\varphi \hat\varphi}^{(n,s)\mathrm{TT}}(\rho ) 
		= - \tilde{h}_{\hat\zeta \hat\zeta}^{(n,s)\mathrm{TT}} (\rho), 
\label{PLUS} \\
	\tilde{h}_{\times}^{(n,s)} (\rho) 
		&= \tilde{h}_{\hat\varphi \hat\zeta }^{(n,s)\mathrm{TT}} (\rho).
\label{CROSS} 
\end{align}

\subsection{Directions of Gravitational wave emission}
Let us consider a set of pairs $(n,s)$ which give the same ratio $s/n$ 
under the conditions  \eqref{OthCon2} and \eqref{positive_kappa}. 
For a stationary rotating string with fixed $l$ and $q$,  
all $(n,s)$-modes in the set are emitted in the same direction 
$\theta_{s/n}$ defined by \eqref{theta}
\cite{footnote}. 
The lowest number of $n$ and corresponding $s$ in the set, say $(n_0, s_0)$ , 
is the fundamental mode of gravitational wave 
emitted to the direction $\theta_{s/n}$. The overtone modes are specified 
by the indices which are multiplications of $(n_0, s_0)$ 
by positive integers larger than 1. 
For example, mode indices of the fundamental mode and the overtone modes 
for each direction are shown in the following table 
in the cases of string with $(l,q)= (0, 1/2)$ and $(1/3, 1/2)$. 

\vspace{0.5cm}
{~\qquad}$(l,q)= (0, 1/2)$ case:
\begin{center}
\begin{tabular}{ccc}
\hline 
$\cos\theta_{s/n}$ &\quad Fundamental mode $(n_0,s_0)$ \quad 
&\quad  Overtone modes $(n,s)$\\
\hline
$0$ & $(2,0)$ & $(4,0), (6,0) \cdots$ \\
$\pm 2/3$ & $(3,\pm 1)$ & $(6,\pm 2), (9,\pm 3) \cdots$\\
$\pm 2/5$ & $(5,\pm 1)$ & $(10,\pm 2), (15,\pm 3) \cdots$\\
$\pm 2/7$ & $(7,\pm 1)$ & $(14,\pm 2), (21,\pm 3) \cdots$\\
$\pm 6/7$ & $(7,\pm 3)$ & $(14,\pm 6), (21,\pm 9) \cdots$\\
$\vdots$ & $\vdots$ & $\vdots$\\
\hline
\end{tabular}
\end{center}
\bigskip
{~\qquad}$(l,q)= (1/3, 1/2)$ case:
\begin{center}
\begin{tabular}{rcc}
\hline 
$\cos\theta_{s/n}$ &\quad Fundamental mode $(n_0,s_0)$ \quad 
&\quad  Overtone modes $(n,s)$\\
\hline
$2/3$ & $(2,1)$ & $(4,2), (6,3) \cdots$\\
$0$ & $(3,0)$ & $(6,0), (9,0) \cdots$ \\
$-1/3$ & $(4,-1)$ & $(8,-2), (12,-3) \cdots$\\
$4/15$ & $(5,1)$ & $(10,2), (15,3) \cdots$\\
$-8/15$ & $(5,-2)$ & $(10,-4), (15,-6) \cdots$\\
$-2/3$ & $(6,-3)$ & $(12,-6), (18,-9) \cdots$\\
$-4/21$ & $(7,-1)$ & $(14,-2), (21,-3) \cdots$\\
$\vdots$ & $\vdots$ & $\vdots$\\
\hline
\end{tabular}
\end{center}
\bigskip

If the direction $\theta_{s/n}$ is fixed, the gravitational wave is 
given by superposition  as
\begin{equation}
	h_{+,\times}^{(s/n)} (t,\rho,\varphi,z) 
		=\sum {}^{'} 
		\left[\exp\left(-in \{ \Omega (t- \cos \theta_{s/n} z) 
			-\varphi \}\right) 
	\tilde{h}_{+,\times}^{(n,s)}(\rho)
	+	(c.c.)\right] ,
\label{htheta}
\end{equation}
where the summation $\Sigma '$ is taken over 
the fundamental mode with frequency $n_0 \Omega$ 
and overtone modes for given $\theta_{s/n}$. 
As will be shown later, the amplitude of the mode with large $n$ is 
highly suppressed, then only several discrete directions are effective for 
gravitational wave emission. 
The discreteness of the directions is analogous to the 
diffraction by gratings. 
This effect comes from the periodic structures of strings. 
Because the stationary rotating strings considered here have 
infinite length along the rotation axis, 
then a distant observer detects gravitational waves coming from 
discrete directions specified by $\theta_{s/n}$. 

In the case of the helical strings $M_l=qN_l/2$ then 
nonvanishing $(n,s)$  modes specified by \eqref{ConHel} 
leads to $\kappa_{ns}^2=0$. 
Therefore, the helical strings do not emit the gravitational wave 
away from the strings.

\subsection{Waveforms} 

The amplitude of gravitational waves behave as 
$1/\sqrt{\rho}$ at the far region because the source string is 
assumed to be infinitely long. 
Then, it is convenient to factorize a nondimensional quantity 
${\mu}/{\sqrt{\Omega \rho}} $ as
\begin{equation}
	h_{+,\times}^{(s/n)} (t,\rho,\varphi,z) 
		= \left( \frac{\mu}{\sqrt{\Omega \rho}} \right) 
			\hat{h}_{+,\times}^{(s/n)} (t,\rho,\varphi,z),
\label{normal} 
\end{equation}
equivalently,
\begin{equation}
	\tilde{h}_{+,\times}^{(n,s)} (\rho) 
		= \left( \frac{\mu}{\sqrt{\Omega \rho}} \right) 
			\hat{\tilde{h}}_{+,\times}^{(n,s)}(\rho) . 
\end{equation}
By this rescaling, the amplitudes of $\hat{h}_{+,\times}^{(s/n)}$ 
are independent of $\mu, \Omega$ and $\rho$ in the far region. 

\begin{figure}[!ht]
\includegraphics[width=14cm]{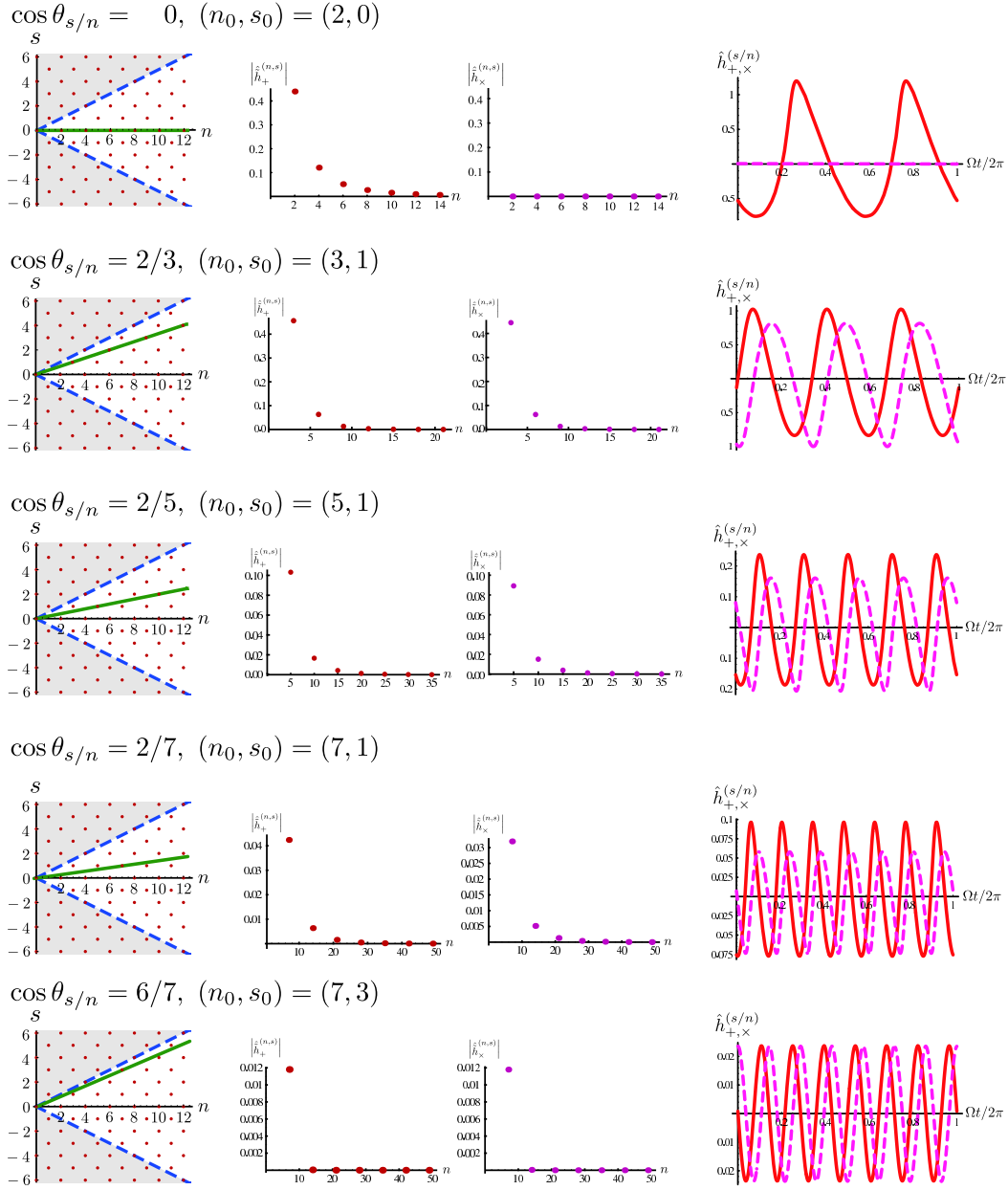} 
\caption{
Waveforms of gravitational waves emitted to 
$\cos\theta_{s/n}=0$, $2/3$, $2/5$, $2/7$, and $6/7$ 
 from a planar string $(l,q)=(0,1/2)$.
In each row, the left panel shows the $(n,s)$ for nonvanishing 
modes.  
The dots on the solid lines correspond to the modes 
which propagate in the direction of $\theta_{s/n}$, 
i.e., fundamental mode and overtone modes for $\theta_{s/n}$. 
The amplitudes $|\hat{\tilde h}_{+,\times}^{(n, s)}|$ 
of the fundamental mode and 
the overtone modes are shown in the middle two panels. 
The right panel shows the waveforms of the plus-mode (solid line) 
and cross modes (broken line). 
}
\label{wfPla}
\end{figure}

\begin{figure}[!ht]
\includegraphics[width=14cm]{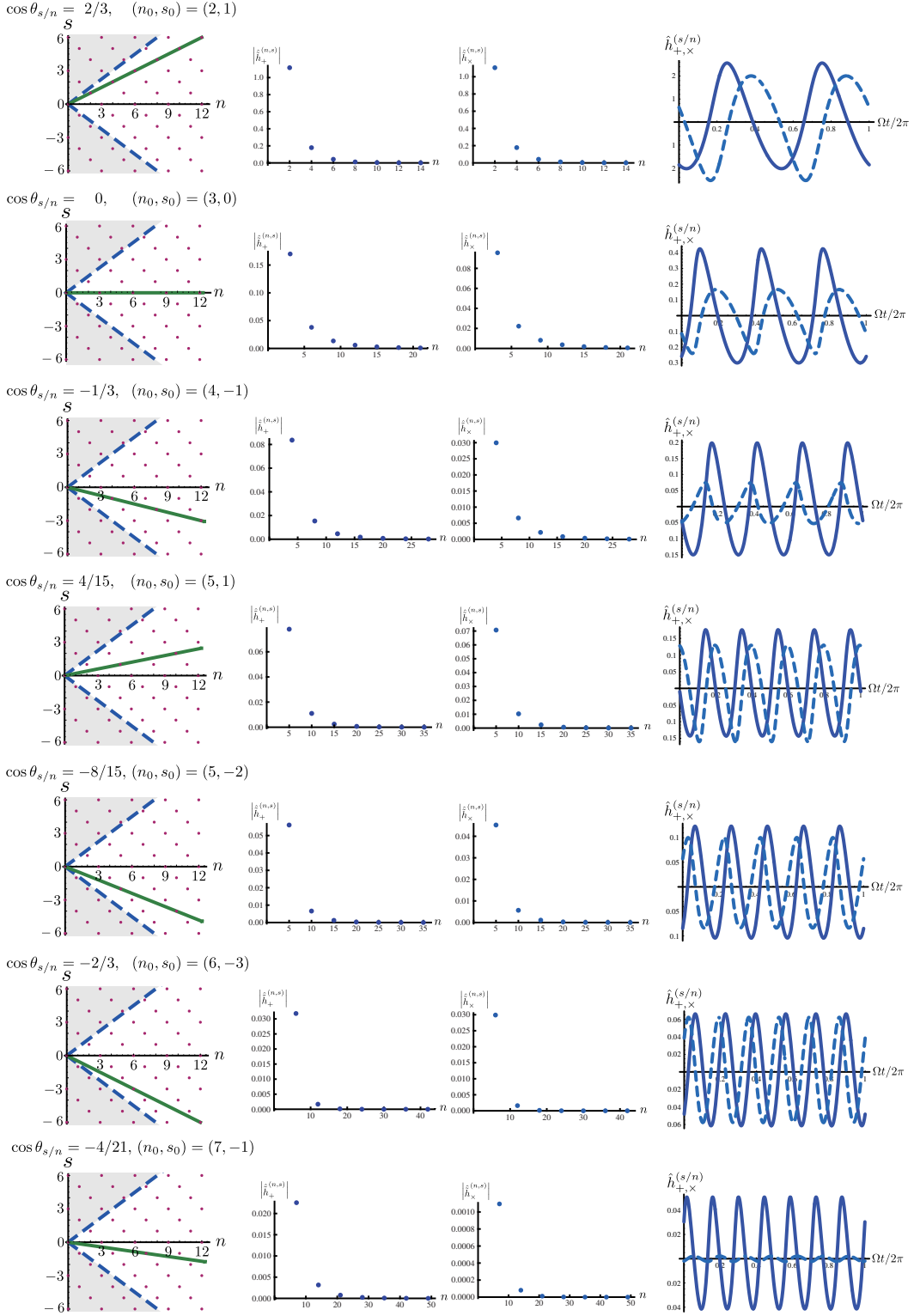} 
\caption{
Waveforms of gravitational waves emitted to the directions 
$\cos\theta_{s/n}=2/3$, $0$, $-1/3$, 
 $4/15$, $-8/15$, $-2/3$, and $-4/21$ 
from the string with $(l, q)=(1/3, 1/2)$.
}
\label{wfTri}
\end{figure}

In Figs.~\ref{wfPla} and \ref{wfTri}, 
we show the waveforms of $\hat{h}_{+,\times}^{(s/n)}$ 
emitted to the direction $\theta_{s/n}$ by the stationary rotating string 
with $(l ,q) =( 0,1/2)$ (planar string),  
and $(l,q)=(1/3,1/2)$, respectively. 
The solid lines and dashed lines in the right figures denote 
the waveform of the plus and cross modes, respectively. 

We can see some characteristic features of the waveforms from 
Figs. \ref{wfPla} and \ref{wfTri}. 
First, 
the waveforms of plus and cross modes are deformed 
from the sine curves of fundamental modes by the overtone modes. 
This is because the magnitudes of the overtone-modes are not negligible. 
\lq Saw-teeth\rq -like shapes appear in the waveforms. 
Secondly 
the amplitude of plus-modes in each direction, $\hat{h}_{+}^{(s/n)}$, 
is determined basically by $n_0$ of the fundamental mode. 
The small $n_0$ gives the large amplitude 
and the large $n_0$ does the small amplitude. 
Third, 
the amplitude of cross modes, in contrast, depends on the direction 
$\theta_{s/n}$. 
The superposition of plus modes and cross modes makes 
\lq almost elliptically polarized waves\rq. The gravitational waves are 
not exactly elliptically polarized because the waves are 
deformed from the sinusoidal form. The \lq ellipticity\rq\ which is given by 
the amplitude ratio of plus and cross modes depend on the direction 
$\theta_{s/n}$. 

In the case of planar strings ($l=0$), 
purely plus-modes are emitted in 
the direction $\theta_{s/n} = \pi/2$, and the cross modes grow as 
$|\theta_{s/n}|$ becomes large. 
When $|\cos\theta_{s/n}|$ approaches to 1, the amplitudes of both modes 
become almost the same, i.e., the waves become the circular polarization. 
In the case of string with $(l,q)=(1/3,1/2)$, 
the amplitude of cross mode is quite small 
in the direction $\cos\theta_{s/n} = -4/21$, and the amplitudes of both modes 
becomes almost the same again as $|\cos\theta_{s/n}|$ approaches to 1.

\section{Traveling Wave Modes } 

We consider, here, the traveling wave modes $h_{\mu \nu}^\mathrm{TW}$ 
given by \eqref{DP}, where $\kappa_{ns}^2=0$. 
Since Green's function is real in this case, 
then $\tilde{h}_{\mu\nu}^{(n,s)} (\rho)$ are real functions of 
$\rho$ with meaningless phase factors; then, 
$h_{\mu \nu}^\mathrm{TW}$ has the form of
\begin{align}
	h_{\mu \nu}^\mathrm{TW}(t,\rho, \varphi, z) 
		=& 2 \sum_{n, s}{}^{'}
	\Bigr\{ \cos \left( n \left\{ \Omega (t - z)-\varphi \right\}\right) 
	\tilde{h}_{\mu\nu}^{(n,s)\mathrm{TT}}(\rho) \Bigr\} 
\nonumber \\
 &+ 2 \sum_{n, -s}{}^{'}
	\Bigr\{ \cos \left( n \left\{ \Omega (t + z)-\varphi \right\}\right) 
 	\tilde{h}_{\mu\nu}^{(n,-s)\mathrm{TT}}(\rho) \Bigr\},
\label{DP2} 
\end{align}
where the integers $n$ and $s$ are required to satisfy the conditions 
\eqref{OthCon2} and \eqref{zero_kappa}. 
From these two conditions, 
$n$ should be a positive integer which satisfies
\begin{equation}
	\frac{ 1-l \pm q}{2} ~n= j ,
\label{TW_cond_n} 
\end{equation}
where $j$ is a positive integer, and $s$ is given by
\begin{equation}
	s=\frac{q N_l}{2} n.
\label{TW_cond_s}
\end{equation}
The condition \eqref{TW_cond_n} means the parameter $q$ 
should be a rational number for appearance of traveling wave modes. 

The summations in \eqref{DP2} are taken over pairs of $(n, s)$ 
and $(n, -s)$ satisfying the conditions \eqref{TW_cond_n} 
and \eqref{TW_cond_s}. 
For example, in the case of $(l,q)=(0,1/2)$, the pairs $(n, \pm s)$ are
$(4,\pm 2), (8,\pm 4), \cdots$.
Nonzero components of $\tilde{h}_{\mu\nu}^{(n,\pm s)\mathrm{TT}}$ are 
\begin{align}
	\tilde{h}_{\rho \rho}^{(n,\pm s)\mathrm{TT}} (\rho)
	=&-\frac{\tilde{h}_{\varphi\varphi}^{(n,s)\mathrm{TT}}(\rho)}{\rho^2}
\cr
	=& -\frac{2 \mu}{q N_l \sigma_\mathrm{p}} 
	\int_0^{N_l \sigma_\mathrm{p}} d\sigma 
	\biggl[ \Bigl\{ 
		G^{n~ \pm s}_{n+2} (\rho,\rho_\mathrm{st}(\sigma))
		\Theta_{\mathrm{T}+}(\rho_\mathrm{st}(\sigma)) 
\cr 
&\qquad 
	+ G^{n~\pm s}_{n-2} (\rho,\rho_\mathrm{st}(\sigma)) 
	\Theta_{\mathrm{T}-} (\rho_\mathrm{st}(\sigma))
	\Bigr\} 
	\exp(-i n\{\pm \Omega q \sigma + \bar{\varphi}(\sigma) \} ) 
	\biggr]
\\
	\frac{\tilde{h}_{\rho \varphi}^{(n,\pm s)\mathrm{TT}} (\rho)}{\rho} 
	= &-\frac{2 \mu i}{q N_l \sigma_\mathrm{p}} 
	\int_0^{N_l \sigma_\mathrm{p}} d\sigma 
	\biggl[ \Bigl\{ 
		G^{n~\pm s}_{n+2} (\rho,\rho_\mathrm{st}(\sigma)) 
		\Theta_{\mathrm{T}+} (\rho_\mathrm{st}(\sigma)) 
\cr
&\qquad 
	- G^{n~\pm s}_{n-2} (\rho,\rho_\mathrm{st}(\sigma)) 
		\Theta_{\mathrm{T}-} (\rho_\mathrm{st}(\sigma))
	\Bigr\} 
	\exp(-i n\{\pm \Omega q \sigma + \bar{\varphi}(\sigma) \} ) 
	\biggr] ,
\label{hPhCpot} 
\end{align}
where Green's functions are given by \eqref{GPL}. 

The modes $h_{\mu \nu}^\mathrm{TW}$ given by \eqref{DP2} consist of 
the superposition of the propagating waves with circular polarization 
in the $(\pm z)$-direction for $\pm s$, respectively. 
We can understand that these modes are obtained by 
the limit $\kappa_{ns}^2 \to 0$ in 
the gravitational wave modes, that is, the direction of wave emission 
in this limit is $\theta_{s/n} = 0, \pi$. 
The metric perturbations of the modes do not propagate off 
the string toward the radial direction. 
These are related to 
the traveling waves discussed in Ref. \cite{Vachaspati}.   

In the helical string cases, from \eqref{ConHel}, 
the first line in  the right hand 
side of \eqref{DP2} vanishes 
and summation in the second line is taken over pairs $(n,-s)$, 
where $n$ is a positive integer and $s$ is given by
\begin{equation}
	s=\frac{(1-l)N_l}{2} n = M_l n.
\end{equation}
There exists only a downward gravitational wave which 
is accompanied with 
the downward string wave \eqref{Helical_Wave}. 

The wave length of each wave propagating along the $z$ axis in the 
traveling wave modes is 
\begin{equation}
	\lambda = \frac{2\pi}{n \Omega} . 
\end{equation}
Then, the condition \eqref{TW_cond_s} for the appearance of 
the traveling wave mode means that the periodicity of 
the stationary rotating string, 
which is given by \eqref{string_period}, 
should be the wavelength of traveling wave times the 
integer $s$, i.e.,
\begin{equation}
	Z_\mathrm{p} = s \lambda . 
\label{wl1p}
\end{equation}
This fact is consistent with the result in Ref.\cite{NII} 
which implies 
that the deformation of the string is caused by the gravitational 
waves propagating on the string.

\section{Summary}

We have studied gravitational perturbations 
around a stationary rotating string in Minkowski spacetime. 
We have solved the linearized Einstein equations with 
the energy-momentum tensor of the string 
by using the one-dimensional Green's function method. 
We have analyzed three long range modes: 
potential mode, 
gravitational wave modes, and traveling wave modes.

\subsection{Potential mode}

The stationary rotating strings produce the logarithmic Newtonian potential 
which is in proportion to $\tilde\mu - \tilde{\cal T}$, 
where $\tilde\mu$ and $\tilde{\cal T}$ denote the effective line density 
and the effective tension of a stationary rotating \lq wiggly\rq\ string 
defined by averaging of the  energy-momentum tensor along its rotation axis. 
The appearance of the Newtonian potential is the result of 
the fact that the effective line density becomes larger than 
the effective tension for rotating strings. 
There also exists angular deficit, $4\pi(\tilde\mu + \tilde{\cal T})$, 
around the string.  

In addition, there are the azimuthal frame-dragging effect 
caused by the angular momentum of the rotating string, 
and the linear frame-dragging along the rotation axis 
caused by the linear momentum $\braket{P}$ of the string 
along the rotation axis. 
The linear frame-dragging disappears if $\braket{P}=0$ in the 
inertial reference frame of observer.  

The helical strings are very special strings. 
Since $\tilde\mu=\tilde{\cal T}$ for the helical strings,
they are not associated with the Newtonian potential, 
and there is an angular deficit with the same amount 
as the straight string case. 
Further, the helical strings cause the linear frame-dragging inevitably 
because there is no inertial reference frame such that $\braket{P}=0$. 
The azimuthal frame-dragging and the linear frame-dragging 
distinguish the helical strings from the straight string.

\subsection{Gravitational wave modes}

The stationary rotating strings 
can emit the gravitational waves in several discrete directions. 
The possible directions for each string are determined by 
the set of parameters $(l, q)$ which specifies the shape of string. 
This property, analogous to the diffraction grating, 
comes from the periodic structure of the strings along the rotation axis. 
The following depend on the directions of gravitational wave emission: 
fundamental frequency, waveforms, amplitude ratio between plus and cross modes,
equivalently, and the ellipticity of elliptic polarization of the waves. 
The waveform of gravitational wave is not the sinusoidal curve 
but a \lq saw-teeth\rq\ like shape. This means that the polarization 
is not exactly elliptical but almost elliptical. 

Since the strings are infinitely long, 
the amplitude of gravitational waves at the large distance 
is proportional to $1/\sqrt{\rho}$. 
Actually, infinite strings are oversimplification.  
But, if the description of the stationary rotating strings is
applicable to a cosmological string in the long range 
comparable to the distance between the string and a observer, 
the amplitude of gravitational waves decreases more gradually than 
the case of point source. 
In this case, it would be possible 
to detect gravitational waves from the stationary rotating strings 
in the cosmological distance (e.g., $\sim 10^3$ Mpc) 
by the present interferometric detectors. 
As the result of the numerical calculations, 
we have obtained the following rough estimation 
of the gravitational wave amplitude:
\begin{equation}
	h_{+,\times} \simeq 
	\mathcal{O} (10^{-14}) 
		\left( \frac{\mu}{10^{-7} } \right) 
		\left( \frac{\Omega/2\pi}{10^3 \mbox{Hz} } \right)^{-1/2} 
		\left( \frac{\rho}{10^3 \mbox{Mpc} } \right)^{-1/2} , 
\end{equation}
where 
we choose $10^{-7}$ as a reference line density of the grand unified theory string, 
$10^3 \mbox{Hz}$ as a reference frequency, 
the most sensitive value 
of the current interferometric detectors 
(TAMA300, LIGO, VIRGO and GEO600), 
and $10^3 \mbox{Mpc} $ as a reference cosmological distance.

\subsection{Traveling wave modes }

As with the special case of gravitational waves, 
traveling waves with the circular polarization propagating 
along the rotating string can appear. 
The strings play the role of wave guide then the amplitude of 
the gravitational wave does not decrease along the string. 
These waves do not propagate off the string toward distant observers, 
but the waves are not confined in the vicinity of the string. 
The amplitude of the traveling waves, described by the power or 
logarithmic function in the radial coordinate, 
gradually decreases as the distance from the string increases. 
Then, even for the distant observer, it would be detectable as the  
gravitational waves propagate parallelly to the strings.

The general stationary rotating strings lose the energy, angular 
momentum, and linear momentum by the gravitational wave emission. 
Then, the strings should evolve by the gravitational radiation. 
If the loss rate of these quantities are small, we can expect 
that the evolution occurs as the transitions 
in the family of the stationary rotating 
strings, approximately. 
What is the final state of strings after the gravitational 
wave emission? 
One would expect that the straight string is the final state. 
But, we should point out that the helical strings are also 
candidates for the final states. 
Because, they do not lose energy, angular momentum and linear momentum by 
the gravitational radiation. They keep the rotation constant 
with traveling waves. 
The study on the final state of the stationary rotating strings 
are now under investigation\cite{WGE}.

\section*{Acknowledgments}
We would like to thank K. Nakao, C.-M. Yoo and S. Saito 
for useful discussions. 
H.I. is supported by Grant-in-Aid for Scientific Research 
Fund of the Ministry of Education, Science and Culture of Japan 
(Grant No. 19540305). 
H.N. is supported by the NSF through grants PHY-0722315, 
PHY-0701566, PHY-0714388, and PHY-0722703.

\appendix
\section{The components of $\Theta^{\mu\nu}$}

The components of $\Theta^{\mu\nu}$ are explicitly expressed in the following:
\begin{equation}\begin{aligned}
	&\Theta^{tt}  = -(1-l^2), 
\quad
	\Theta^{t\rho}  = \frac{l\, \Omega }{2 \rho }
	(\rho_\mathrm{max}^2 -\rho_\mathrm{min}^2) 
		\sin\left(2|\Omega|\sigma \right),  
\quad
	\rho \, \Theta^{t\varphi}  
		= -\frac{\Omega^2 \rho^2 - l^2}{\Omega \rho }, 
\\
	&\Theta^{tz}  = ql \mathrm{sign}(\Omega), 
\quad
	\Theta^{\rho\rho}  
		= \frac{\Omega^2 }{4\rho^2 } 
 	(\rho_\mathrm{max}^2 -\rho_\mathrm{min}^2)^2
		\sin^2\left(2|\Omega|\sigma\right), 
\\
	&\rho\, \Theta^{\rho\varphi}  
		= \frac{l}{2\rho^2}
 		(\rho_\mathrm{max}^2 -\rho_\mathrm{min}^2)
		\sin\left(2|\Omega|\sigma\right),  
\quad
	\Theta^{\rho z}  
		= \frac{q|\Omega|}{2\rho}(\rho_\mathrm{max}^2 -\rho_\mathrm{min}^2)
 			\sin\left(2|\Omega|\sigma\right),  
\\
	&\rho^2  \,\Theta^{\varphi \varphi}  
		= -\Omega^2 \rho^2
		\left( 1- \frac{l^2}{\Omega^4 \rho^4 } \right), 
\quad
	\rho  \, \Theta^{\varphi z}  
		= \frac{l q}{|\Omega| \rho },  
\quad
	\Theta^{zz}  = q^2. 
\label{Thetamunu} 
\end{aligned}\end{equation}

In these expressions, $\rho=\rho(\sigma)$ is given by \eqref{SRSsol}. 




\begin{references}
\bibitem{footnote}
{
When the parameter $l$ of the stationary rotating string is irrational, 
the string does not emit any gravitational wave to an distant observer. 
This would be caused by the phase cancellation of the waves from the 
infinitely long string. 
}
\end{references}
\end{document}